\documentclass[reprint,amsmath,amssymb, aps, prl,superscriptaddress,email]{revtex4-2}

\usepackage{siunitx}
\usepackage{physics}
\usepackage{graphicx}
\usepackage{xcolor}
\usepackage{bm}
\usepackage{natbib}
\usepackage{blindtext}
\usepackage{hyperref}
\usepackage{fontawesome}
\usepackage{subfiles} % Best loaded last in the preamble

\begin{document}

\title{Avalanche terahertz photon detection in a Rydberg tweezer array}

\author{Chris Nill}
\affiliation{Institut f\"ur Theoretische Physik, Universit\"at Tübingen, Auf der Morgenstelle 14, 72076 T\"ubingen, Germany}
\affiliation{Institute for Applied Physics, University of Bonn, Wegelerstraße 8, 53115 Bonn, Germany}
\author{Albert Cabot}
\affiliation{Institut f\"ur Theoretische Physik, Universit\"at Tübingen, Auf der Morgenstelle 14, 72076 T\"ubingen, Germany}
\author{Arno Trautmann}
\affiliation{Physikalisches Institut, Universit\"{a}t T\"{u}bingen, Auf der Morgenstelle 14, 72076 T\"{u}bingen, Germany}
\author{Christian Groß}
\affiliation{Physikalisches Institut, Universit\"{a}t T\"{u}bingen, Auf der Morgenstelle 14, 72076 T\"{u}bingen, Germany}
\author{Igor Lesanovsky}
\affiliation{Institut f\"ur Theoretische Physik, Universit\"at Tübingen, Auf der Morgenstelle 14, 72076 T\"ubingen, Germany}
\affiliation{School of Physics and Astronomy and Centre for the Mathematics and Theoretical Physics of Quantum Non-Equilibrium Systems, The University of Nottingham, Nottingham, NG7 2RD, United Kingdom}

\begin{abstract}
	We propose a protocol for the amplified detection of low-intensity terahertz radiation using Rydberg tweezer arrays. The protocol offers single photon sensitivity together with a low dark count rate. It is split into two phases: during a sensing phase, it harnesses strong terahertz-range transitions between highly excited Rydberg states to capture individual terahertz photons. During an amplification phase it exploits the Rydberg facilitation mechanism which converts a single terahertz photon into a substantial signal of Rydberg excitations. We discuss a concrete realization based on realistic atomic interaction parameters, develop a comprehensive theoretical model that incorporates the motion of trapped atoms and study the many-body dynamics using tensor network methods.
\end{abstract}

\maketitle

\textbf{Introduction --- } 
When an atom is excited to a high-lying Rydberg state the valence electron and the remaining positively charged core form a giant electric dipole \cite{Saffman2010}. Rydberg atoms are thus highly susceptible to electric fields and can find applications in a variety of sensors~\cite{Adams2019, Simons2021}, for example, for detecting small dc field variations \cite{Osterwalder1999, Mohapatra2008}. Another important property of Rydberg atoms is that their spectrum features strong dipole-transitions across a wide range of frequencies, including the terahertz (THz) regime  \cite{Sedlacek2012, Wade2017}. This property, together with their large electric dipole moment, permits the realization of THz sensors offering spatial and temporal resolution for the detection of classical fields \cite{Downes2020}.

In this work we propose and theoretically investigate a protocol which allows for the \textit{amplified} detection of THz radiation at the single THz photon level utilizing additionally the strong state-dependent inter-atomic interactions between Rydberg atoms. These interactions have already been used for enhanced metrological protocols~\cite{Wade2018, Ding2022, Eckner2023, Hines2023}. Our detector is based on a Rydberg tweezer array in which ground state atoms are laser-excited to a Rydberg state. Absorption of a THz photon triggers the transition of an excited atom to a second Rydberg state. Carefully chosen inter- and intra-state interactions then initiate a facilitation dynamics resulting in an avalanche amplification of an absorbed THz photon. We characterize the detector and discuss limitations and imperfections, e.g., resulting from the coupling of the electronic dynamics to lattice vibrations. Beyond providing single photon sensitivity in the THz regime the proposed detector offers a low dark count rate, and therefore may find applications as a sensing device in dark matter searches~\cite{Yamamoto1999,Graham2023}.

\begin{figure*}
	\centering
	\includegraphics[width=\linewidth]{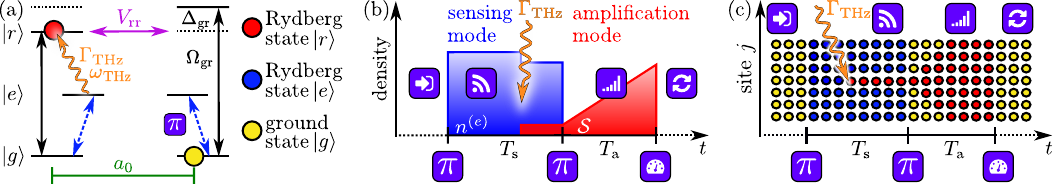}
	\caption{\textbf{Terahertz sensor using a Rydberg tweezer array.}
	(a) Atoms (ground state $\ket{g}$, Rydberg states $\ket{e}$ and $\ket{r}$) with spacing $a_0$ interact with a density-density interaction $V_\mathrm{rr}$ when in Rydberg state $\ket{r}$.
	An off-resonant laser with Rabi frequency $\Omega_\mathrm{gr}$ and detuning $\Delta_\mathrm{gr}$  $(\Omega_\mathrm{gr}\ll|\Delta_\mathrm{gr}|)$ drives the ground-to-Rydberg state $\ket{r}$ transition.
	An additional laser is used for the $\pi$-pulse.
	Terahertz absorption at a frequency $\omega_\mathrm{THz}$ takes place between the Rydberg states with rate $\Gamma_\mathrm {THz}$.
	(b) Detection protocol:
    \faicon{sign-in}: All atoms are initialized in state $\ket{g}$.
    The $\pi$-pulse excites all atoms in state $\ket{e}$.
    \faicon{rss}: During the sensing time $T_s$, atoms can transition to $\ket{r}$ by absorbing a THz photon.
    After applying a second $\pi$-pulse, all atoms from state $\ket{e}$ transition to $\ket{g}$.
    \faicon{signal}: During the amplification phase $T_a$ the number of Rydberg atoms increases due to the facilitated avalanche.
    \faicon{dashboard}: The final measurement of Rydberg atoms yields the amplified signal.
    \faicon{refresh}: The process can be restarted from the beginning.
    (c) Sketch of the spatially resolved evolution of the atomic states. The avalanche is triggered with the second $\pi$-pulse.
	}
	\label{fig:cartoon}
\end{figure*}

\textbf{Atomic model ---} To illustrate the basic idea behind the detector we consider a one-dimensional open-boundary chain of $N$ Rydberg atoms. Neighboring atoms are positioned at an interatomic distance $a_0$ as depicted in Fig. \ref{fig:cartoon}a. Such setting and higher-dimensional generalizations of it can be realized with the help of optical tweezer arrays \cite{Browaeys2020}.
The Rydberg atoms are modeled as three-level systems consisting of a ground state $\ket{g}$ and two Rydberg states $\ket{e}$ and $\ket{r}$.
The transition frequency, $\omega_\mathrm{THz}$, between the two Rydberg states can be chosen across a wide frequency range, including THz. Two neighboring atoms in the Rydberg state $\ket{r}$ interact with a density-density interaction $V_\mathrm{rr}$. An off-resonant laser with a Rabi frequency $\Omega_\mathrm{gr}$ and a detuning $\Delta_\mathrm{gr}$ ($\Omega_\mathrm{gr}\ll|\Delta_\mathrm{gr}|$) couples the state $\ket{g}$ and $\ket{r}$. The laser detuning is set such that it cancels out the interaction between two consecutive atoms in the $\ket{r}$ state, i.e. $\Delta_\mathrm{gr} + V_\mathrm{rr} = 0$. This is the so-called facilitation condition, which has been experimentally and theoretically explored in a various settings \cite{Amthor2010,Ates2007,Mattioli2015,Valado2016,Letscher2017,Su2017,Gambetta2020a,Festa2022,Liu2022}. It ensures that the excitation of an atom to the $\ket{r}$ state is strongly enhanced when it is located next to an atom already in state $\ket{r}$. This process is at the heart of the conversion of a THz photon into a detectable avalanche of Rydberg atoms.

\textbf{Dynamical THz detection protocol ---} The THz detection protocol is depicted schematically in Fig. \ref{fig:cartoon}b,c. It consists of four steps: preparation of the initial state, the sensing mode, the amplification mode and, finally, the measurement.
In the first step, the tweezer array is loaded with atoms in their ground state. The state of the many-body system is then $\ket{\Psi_\mathrm{g}}=\otimes_j \ket{g}_j$, where the sites are labeled by the index $j$. 
Next the \textit{sensing mode} is initialized by a $\pi$-pulse to the $\ket{e}$-state, such that the many-body wave function is $\ket{\Psi_\mathrm{s}}=\otimes_j \ket{e}_j$.

For a time interval of length $T_\mathrm{s}$ the atoms can absorb a THz photon triggering the transition to the second Rydberg state $\ket{r}$. We assume that the rate of absorption, $\Gamma_\mathrm{THz}$, is sufficiently small such that at most one photon is absorbed in the sensing time window: $\Gamma_\mathrm{THz}T_\mathrm{s}\ll 1$. This is no fundamental limitation as amplification is also possible for multiple excitations. Assuming absorption of a photon at site $k$, the state of the system becomes 
$\ket{\Psi_\mathrm{er}}=\bigotimes_{j\neq k} \ket{e}_j\ket{r}_k$.
Note, that the wavelength of THz radiation is typically much larger than the characteristic interatomic distance \cite{Gross1982,Hao2021, Suarez2022}, which actually leads to a collective excitation in state $\ket{r}$. This case will be considered further below.
After the sensing interval a $\pi$-pulse de-excites the atoms in state $\ket{e}$ to the ground state $\ket{g}$, resulting in the state $\ket{\Psi_\mathrm{gr}}=\bigotimes_{j\neq k} \ket{g}_j\ket{r}_k$.
Here we require the interaction between two atoms in state $\ket{e}$, $V_\mathrm{ee}$, and the dipolar exchange interaction, $V_\mathrm{er}$, to be sufficiently small. Below, we will discuss how these conditions can indeed be met in a realistic setting.

Next is the \textit{amplification mode}, which lasts for a time $T_a$, as shown in Fig. \ref{fig:cartoon}b,c. Here the dynamics is described by the Hamiltonian ($\hbar=1$)
\begin{align}
    H_\mathrm{a}&=\Omega_\mathrm{gr} \sum_j \left(\dyad{r}{g}_j + \mathrm{h.c.} \right)+\Delta_\mathrm{gr} \sum_j n^{(\mathrm{r})}_j\nonumber\\
    &+ V_\mathrm{rr}\sum_j n^{(\mathrm{r})}_j n^{(\mathrm{r})}_{j+1},
    \label{eq:Hamiltonian-amplification}
\end{align}
where the laser detuning $\Delta_\mathrm{gr}$ is chosen to match the facilitation condition, i.e. $\Delta_\mathrm{gr}+V_\mathrm{rr}=0$. Moreover, we chose the detuning to be much larger than the Rabi frequency, $\Omega_\mathrm{gr}\ll|\Delta_\mathrm{gr}|$. This condition ensures that predominantly facilitated excitations take place. Off-resonant excitations limit the dark count rate, which can be minimized by an optimal $|\Delta_\mathrm{gr}|$. The facilitation process leads to a large number of Rydberg atoms conditioned on the presence of one THz-excited atom in state $\ket{r}$ (see Fig. \ref{fig:cartoon}c). 

After the amplification time, the number of Rydberg atoms in the state $\ket{r}$ is measured, whose average is given by the signal
\begin{equation}
    \mathcal{S}=\sum_j \mathcal{S}_j=\sum_j \bra{\Psi_\mathrm{gr}}e^{iH_\mathrm{a} T_\mathrm{a}}n^{(\mathrm{r})}_j e^{-iH_\mathrm{a} T_\mathrm{a}}\ket{\Psi_\mathrm{gr}}.
    \label{eq:signal}
\end{equation}
Here $\mathcal{S}_j$ is the spatially resolved signal, i.e., the probability of having a Rydberg atom in state $\ket{r}$ on site $j$. Ideally, the signal $\mathcal{S}$ is proportional to the total number of atoms. Our analysis is only strictly valid for times $T_\mathrm{a}$ that are much smaller than the lifetime of the Rydberg atoms. However, in a typical experimental setting each decayed Rydberg atom is lost from the system. Unless, the loss takes place at the facilitation front where it interrupts the avalanche, absent atoms in the bulk can simply be included in the Rydberg count. In the following we discuss the quantum dynamics of this part of the protocol, taking into account interatomic forces and the fact that the initial THz absorption is collective. We provide experimental parameters below.

\begin{figure}
	\centering
	\includegraphics[width=\linewidth]{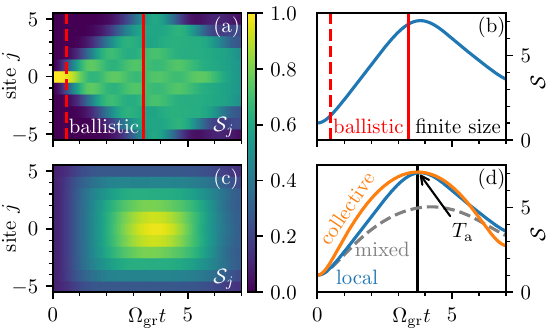}
	\caption{\textbf{Time evolution of the signal after absorption of a THz photon.} (a,c) Time evolution of the spatially resolved signal $\mathcal{S}_j$ in a 1D atom chain with open boundary conditions.
	In panel (a) site $k=0$ absorbed a local THz photon leading to the initial state $\ket{\Psi_\mathrm{gr}}$. Panel (c) shows the collective absorption case where the initial state is $\ket{\Psi_\mathrm{gr}^\mathrm{c}}$.
	We set $V_\mathrm{rr}=-\Delta_\mathrm{gr}=500\Omega_\mathrm{gr}$.
	(b) Time evolution of the signal $\mathcal{S}$. After a first initial phase, the signal increases linearly (ballistic expansion of Rydberg excitation cluster) until the whole 1D chain is excited. What follows are finite size effects.
	Consequently, the optimal measurement time is directly at the end of the ballistic expansion phase where $\mathcal{S}$ is maximal.
	(d) Time evolution of the signal $\mathcal{S}$ after collective absorption (orange line), local absorption at site $k=0$ (blue line) and local absorption, averaged with equal weight over all possible initial positions (gray dashed).
	We mark by $T_\text{a}$ the time of optimal amplification.
	The simulation was done integrating the corresponding Schr\"odinger equation \cite{Johansson2012,Johansson2013}.
	}
	\label{fig:combined-facilitation}
    \end{figure}

\textbf{Many-body dynamics during amplification mode ---}
The facilitation excitation dynamics taking place during amplification is illustrated in Fig. \ref{fig:combined-facilitation}a, where we present the spatially resolved signal $\mathcal{S}_j$ as a function of time $t$, starting from the initial state $\ket{\Psi_\mathrm{gr}}$. For the case shown the THz photon was absorbed by the central atom, located at site $k=0$. In Fig. \ref{fig:combined-facilitation}b we show the time evolution for the total signal $\mathcal{S}$, see Eq. (\ref{eq:signal}).

The dynamics is characterized by three stages, which are delimited by the vertical red dashed and solid lines in Fig. \ref{fig:combined-facilitation}b.
In the first stage we observe a quadratic increase of the signal $\mathcal{S}\propto(\Omega_\mathrm{gr}t)^2$. In the second stage a ballistic expansion is established (see region between red dashed and solid lines in Fig. \ref{fig:combined-facilitation}b). Here, the already facilitated Rydberg atoms excite their neighbors leading to the creation of clusters of consecutive Rydberg excitations. This cluster grows from the boundaries (the facilitation front) as the de-excitation of Rydberg atoms within the bulk is off-resonant: atoms in state $\ket{r}$ experience an energy-shift, which is $2V_\mathrm{rr}$, since they are interacting with their left and right neighbor. During this ballistic expansion the number of Rydberg atoms grows approximately linearly in time and so does the signal: $\mathcal{S}\propto \Omega_\mathrm{gr}t$, leading to higher and higher amplification \cite{supplement}.
A third stage follows in which the dynamics is governed by finite size effects and the signal $\mathcal{S}$ starts to reduce once the edges of the Rydberg cluster hit the boundaries of the lattice. This implies that there is an optimal value for the amplification time $T_\text{a}$, which is proportional to $N/\Omega_\mathrm{gr}$. Note, that this is indeed a quantity that can be optimized: the absorption of a THz photon may take place at a random time. However, the starting point of the amplification mode is precisely known. Note furthermore, that the inclusion of dephasing yields a saturation of the signal at $\mathcal{S}\approx N/2$ in the long time limit \cite{supplement}.

So far, we have assumed that the absorption of a THz photon takes place at a specific site $k$ of the atom chain.
However, terahertz radiation has a much longer wavelength than the typical interatomic distances $a_0$ in tweezer arrays.
Absorption of the THz photon is then described by the collective jump operator $L=\sqrt{\Gamma_\text{THz}}\sum_j \dyad{r}{e}_j$  \cite{Gross1982,Bettles2016,Hao2021}.
This creates the coherent superposition state $\ket{\Psi^\mathrm{c}_\mathrm{er}}=\frac{1}{\sqrt{N}}\sum_k \bigotimes_{j\neq k}\ket{e}_j\ket{r}_k\propto L\ket{\Psi_\text{s}}$,
where the excitation in the Rydberg state $\ket{r}$ is collective, i.e. shared among the entire ensemble (see \cite{supplement} for more details). At the end of the sensing phase this state is de-excited to
$\ket{\Psi^\mathrm{c}_\mathrm{gr}}=\frac{1}{\sqrt{N}}\sum_k \bigotimes_{j\neq k}\ket{g}_j\ket{r}_k$. Starting the amplification mode leads to the signal shown in Fig. \ref{fig:combined-facilitation}c,d, which increases notably faster than that of a local excitation. This acceleration is actually a coherent effect owed to the collective nature of the state $\ket{\Psi^\mathrm{c}_\mathrm{gr}}$. To see this, we show for comparison the signal for a mixed state, corresponding to the incoherent equal weight average over all possible initial positions of the atom in the $\ket{r}$-state. This signal is lower than the one corresponding to the excitation of the central atom (Fig. \ref{fig:combined-facilitation}a), which is expected since the latter case produces more facilitated atoms than an initial excitation close to the boundary. 

\textbf{Experimental considerations ---}
As indicated previously, the implementation of the detector protocol requires specifically chosen interactions: interactions between Rydberg $\ket{r}$-states shall be strongest, while interactions between atoms in the $\ket{e}$-state and crossed interactions between atoms in the $\ket{e}$- and $\ket{r}$-states shall be small compared to the relevant laser Rabi frequency. Moreover, the detector is spectrally sensitive solely near the frequency $\omega_\mathrm{THz}$, which, however, can be tuned  over a wide range. We illustrate this in the following by considering two exemplary cases for the element $^{39}$K.

\begin{figure}[t]
	\centering
	\includegraphics[width=\linewidth]{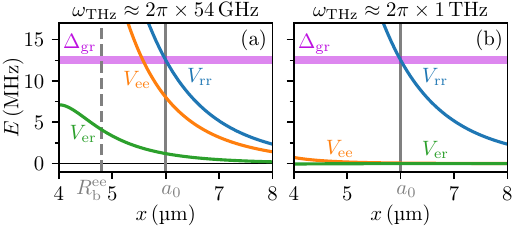}
	\caption{\textbf{Interactions between Rydberg states of $^{39}\mathrm{K}$.}
	(a) Pair potentials $V_\mathrm{rr}$ (blue), $V_\mathrm{ee}$ (orange), and $V_\mathrm{er}$ (green) as a function of the atomic separation $x$ for $\ket{r} = 70P_{1/2}$ and $\ket{e}= 68S_{1/2}$, with $\omega_\mathrm{THz} \approx 2\pi\times\SI{54}{\giga\hertz}$.
	The horizontal purple lines indicate the laser detuning $\Delta_\mathrm{gr}$ for facilitating the $\ket g \to \ket r$-transition at a distance of $a_0=\SI{6}{\micro\meter}$ (grey vertical line).
	The dashed grey vertical line in shows the blockade radius $R_\mathrm{b}^\mathrm{ee}$ for the $\ket{g} \to \ket {e}$ excitation.
	(b) Pair potentials for $\ket{r} = 70P_{1/2}$ and $\ket{e} = 45S_{1/2}$. The color scheme is the same as in (a). $V_\mathrm{er}$ is not visible on this scale. In both cases we assumed a magnetic offset field of $\SI{5}{G}$. Calculations have been performed with the ``pairinteraction'' software~\cite{Weber2017}.
	}
	\label{fig:pairpotentials}
\end{figure}

We choose the Rydberg state $\ket r = 70P_{1/2}$ and two different sensing states, i.e. states from which the atom is excited into $\ket r$ upon absorption of the photon: (a) $\ket e_{(a)} = 68S_{1/2}$ and (b) $\ket e_{(b)} = 45S_{1/2}$. Using state (a) the transition energy is $\SI{54}{GHz}$, a convenient frequency for a laboratory microwave source for demonstration of the scheme, while for state (b) the transition is in the THz regime, with about $\SI{1}{THz}$. The corresponding interaction potentials are shown in Fig. \ref{fig:pairpotentials}. Choosing an interatomic distance $a_0 = \SI6{\micro m}$, in scenario (a) the interaction energy is $V_\mathrm{rr} \approx \SI{12.5}{MHz}$, $V_\mathrm{ee} \approx \SI{9}{MHz}$, and $V_\mathrm{er} \approx \SI{1}{MHz}$. With a Rabi frequency of $\Omega_\mathrm{ge} = 2\pi\times\SI{30}{MHz}$, the Rydberg blockade at this distance can be broken, effectively allowing to neglect the interaction in $\ket e$. Moreover, as the interaction between $\ket r$ and $\ket e$ is much smaller than $V_\mathrm{ee}$, an atom excited to $\ket e$ will not affect the remaining atoms in $\ket r$. Thus both excitation and de-excitation can approximately be treated in the limit of non-interacting states as was assumed previously. 

The potential $V_\mathrm{rr}$ (and thus the laser detuning: $\Delta_\mathrm{gr}=-V_\mathrm{rr}$) is large enough to suppress off-resonant scattering during sensing: choosing a Rabi frequency of $\Omega_\mathrm{gr} = 2\pi\times\SI{0.2}{MHz}$ and a detuning $\Delta_\mathrm{gr} = \SI{12.5}{MHz}$ results in a dark count rate of \SI{0.33}{\per \second}. In a cryogenic environment at \SI1K this is further reduced to \SI{0.05}{\per \second} due to lower influence of black-body radiation on broadening of the Rydberg state's absorption. For an array of $11$~atoms, these parameters give an optimal amplification time of about \SI{25}{\micro s}, which leaves about \SI{50}{\micro s} for the sensing before significant atom loss from the tweezer array. The lifetimes of the Rydberg states are sufficiently long, with $\tau_e = \SI{193}{\micro s}\ (\SI{1.2}{ms}), \tau_r = \SI{129}{\micro s}\  (\SI{330}{\micro s})$, and $\tau_{45S} = \SI{46}{\micro s}\  (\SI{91}{\micro s})$ at \SI{330}{K} (\SI{1}{K}). 
Assuming that Rydberg atoms are not lost from the system, e.g. by using state-independent trapping, the detector dead time is thus given by the readout time of about \SI{10}{ms}, after which the next sensing cycle can start (the cycle rate is thus $\sim100$ Hz). 

Case (b) refers to a frequency in the range of \mbox{$\omega_\mathrm{THz}\approx \SI{1}{THz}$.} Here, the potentials $V_\mathrm{ee}$ and $V_\mathrm{er}$ are even weaker than in case (a), as the relevant dipole transition matrix elements are diminished by a small wave function overlap. Therefore, all required conditions are naturally met. The  weak interaction of the $\ket e$ states also allows to reduce the lattice spacing $a_0$, which increases $\Delta_\mathrm{gr}$ and suppresses the dark count rate further.\\

\textbf{Facilitation dynamics and atomic motion ---}
The faciliation mechanism is highly sensitive to the distance between neighboring atoms \cite{Marcuzzi2017,Ostmann2019,Wintermantel2021}. We therefore investigate in the following the impact of atomic motion within the tweezer traps. For simplicity, we assume the atoms to be trapped in a state-independent potential \cite{Anderson2011,Li2013, Goldschmidt2015, Wilson2022}.
Traps are modeled as harmonic oscillators with frequency $\nu$ and bosonic lowering and rising operators $a_j$ and $a_j^\dagger$ acting at site $j$.
This is valid as long as the atoms are cooled near the motional ground state $(\nu\gg kT)$ with thermal energy $kT$ \cite{Kaufman2012, Thompson2013, Cooper2018, Norcia2018, Saskin2019, Lorenz2021, Hoelzl2023}.
Coupling between the Rydberg facilitation dynamics and the vibrational motion is caused by the dependence of the interaction potential $V_\mathrm{rr}$ on the interatomic separation $x$: $V_\mathrm{rr}\rightarrow V_\mathrm{rr}(x)$. The Hamiltonian $H_\mathrm{a}$ [Eq. (\ref{eq:Hamiltonian-amplification})], valid during the amplification mode, therefore changes to \cite{Magoni2022,Magoni2023a}
\begin{equation}
    H_\mathrm{a}'=
    H_\mathrm{a}
    +\nu\sum_j a_j^\dagger a_j
    +\pdv{V_\mathrm{rr}(x)}{x}\Biggr|_{x=a_0}\!\!\!\sum_j n^{(\mathrm{r})}_j n^{(\mathrm{r})}_{j+1}\delta x^{(j,j+1)}.
\end{equation}
Note, that we considered here only small (first order) displacements of the atoms from their equilibrium positions. These displacements are represented by the operator $\delta x^{(j,j+1)}= 1/\sqrt{2m\nu} (a_j+a_j^\dagger-a_{j+1}-a_{j+1}^\dagger)$, where $m$ is the atom mass.
The coupling strength between the vibrational degree of freedom and the Rydberg state of the atoms is then given by $\kappa=1/\sqrt{2m\nu}\ \partial_x V_\mathrm{rr}(x)|_{x=a_0}$ \cite{Magoni2022}.
To simulate the dynamics of the ensuing spin-boson Hamiltonian we resort to the time-evolving block decimation algorithm (TEBD) \cite{Vidal2003,Vidal2004,Orus2014,Paeckel2019,Cirac2021,Gray2018}, and truncate the Fock space of the harmonic oscillators at a maximum of $7$ phonons.

\begin{figure}[t!]
    \centering
    \includegraphics[width=\linewidth]{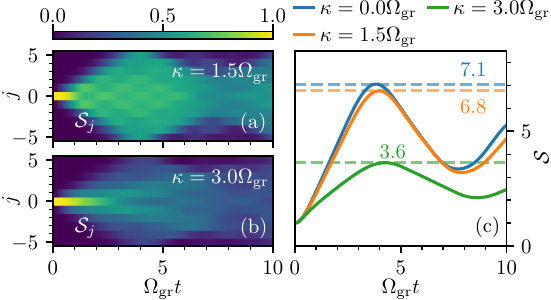}
    \caption{\textbf{Impact of atomic motion.} (a,b) Facilitation dynamics, visible in the spatially resolved $\mathcal{S}_j$, for two values of the electronic-vibrational coupling strength $\kappa$. Increasing $\kappa$ generally decreases the signal. (c) The maximally achievable amplification (dashed lines and numerical values) decreases as $\kappa$ increases.
    Even for moderate coupling strength $(\kappa=3\Omega_\mathrm{gr})$ the amplification of a THz-excited Rydberg atom in the $\ket{r}$-state is clearly observed. In the simulations the trap frequency is $\nu=8\Omega_\mathrm{gr}$.
    }
    \label{fig:vibrations}
\end{figure}

Figure \ref{fig:vibrations}a,b show the dynamics during amplification mode. While Rydberg excitations, triggered by the absorption of a THz, are still spreading, we observe in Fig. \ref{fig:vibrations}c that the maximally achievable amplification generally decreases when the coupling strength $\kappa$ between the vibrational and electronic degrees of freedom is increased.
Nevertheless, for $\kappa=1.5\Omega_\mathrm{gr}$ there is only a minimal change compared to the uncoupled case ($\kappa=0$), and even for stronger couplings significant amplification is possible. Thus, robust trapping with $\kappa\ll\nu$ is certainly advantageous, but efficient amplification is possible when vibrational atomic motion is present.

\textbf{Conclusions and future directions --- }
We have discussed a protocol for a THz photon avalanche detector that combines the tunable frequency range of transitions among Rydberg states with facilitated Rydberg excitation. The detector offers single photon sensitivity together with a low dark count rate and fast operation cycle. One possibility to further enhance the sensitivity of the detector is to use correlated initial states: rather than initializing all the atoms in the $\ket{e}$-state one may think of preparing them in a Dicke state containing in the manifold of $\ket{e}$,$\ket{r}$-states. This would allow to collectively (superradiantly) enhance the absorption of THz photons. However, implementing such protocol requires switchable interactions between $\ket{r}$-states and a generalization of the avalanche dynamics into the many-body regime, i.e. where multiple excitations are initially present. Beyond the microscopically controlled optical tweezer arrays, as discussed here, the protocol is expected to also work in disordered gases as long as Doppler-broadening is significantly smaller than $\Delta_\text{gr}$. The large number of atoms in these gases enhances the THz absorption probability. At the same time single photon sensitivity is maintained due to the large signal produced by the avalanche. 

\acknowledgments
\textbf{Acknowledgements --- }
We acknowledge funding from the Deutsche Forschungsgemeinschaft within SPP 1929 GiRyd (Grant No. 428276754: LE3522/1 and GR4741/5), a Heisenberg professorship to C.G. (GR4741/3) and the research units FOR5413 (Grant No. 465199066) and FOR5522 (Grant No. 499180199). We also acknowledge funding from the Horizon Europe programme HORIZON-CL4-2022-QUANTUM-02-SGA via the project 101113690 (PASQuanS2.1), the Baden-W\"urttemberg Stiftung through Project No.~BWST\_ISF2019-23, the Alfried Krupp von Bohlen and Halbach foundation and the state of Baden-Württemberg through bwHPC grant no INST 40/575-1 FUGG (JUSTUS 2 cluster).

\bibliography{biblio}

\begin{thebibliography}{55}%
\makeatletter
\providecommand \@ifxundefined [1]{%
 \@ifx{#1\undefined}
}%
\providecommand \@ifnum [1]{%
 \ifnum #1\expandafter \@firstoftwo
 \else \expandafter \@secondoftwo
 \fi
}%
\providecommand \@ifx [1]{%
 \ifx #1\expandafter \@firstoftwo
 \else \expandafter \@secondoftwo
 \fi
}%
\providecommand \natexlab [1]{#1}%
\providecommand \enquote  [1]{``#1''}%
\providecommand \bibnamefont  [1]{#1}%
\providecommand \bibfnamefont [1]{#1}%
\providecommand \citenamefont [1]{#1}%
\providecommand \href@noop [0]{\@secondoftwo}%
\providecommand \href [0]{\begingroup \@sanitize@url \@href}%
\providecommand \@href[1]{\@@startlink{#1}\@@href}%
\providecommand \@@href[1]{\endgroup#1\@@endlink}%
\providecommand \@sanitize@url [0]{\catcode `\\12\catcode `\$12\catcode
  `\&12\catcode `\#12\catcode `\^12\catcode `\_12\catcode `\%12\relax}%
\providecommand \@@startlink[1]{}%
\providecommand \@@endlink[0]{}%
\providecommand \url  [0]{\begingroup\@sanitize@url \@url }%
\providecommand \@url [1]{\endgroup\@href {#1}{\urlprefix }}%
\providecommand \urlprefix  [0]{URL }%
\providecommand \Eprint [0]{\href }%
\providecommand \doibase [0]{https://doi.org/}%
\providecommand \selectlanguage [0]{\@gobble}%
\providecommand \bibinfo  [0]{\@secondoftwo}%
\providecommand \bibfield  [0]{\@secondoftwo}%
\providecommand \translation [1]{[#1]}%
\providecommand \BibitemOpen [0]{}%
\providecommand \bibitemStop [0]{}%
\providecommand \bibitemNoStop [0]{.\EOS\space}%
\providecommand \EOS [0]{\spacefactor3000\relax}%
\providecommand \BibitemShut  [1]{\csname bibitem#1\endcsname}%
\let\auto@bib@innerbib\@empty
%</preamble>
\bibitem [{\citenamefont {Saffman}\ \emph {et~al.}(2010)\citenamefont
  {Saffman}, \citenamefont {Walker},\ and\ \citenamefont
  {M{\o}lmer}}]{Saffman2010}%
  \BibitemOpen
  \bibfield  {author} {\bibinfo {author} {\bibfnamefont {M.}~\bibnamefont
  {Saffman}}, \bibinfo {author} {\bibfnamefont {T.~G.}\ \bibnamefont
  {Walker}},\ and\ \bibinfo {author} {\bibfnamefont {K.}~\bibnamefont
  {M{\o}lmer}},\ }\bibfield  {title} {\bibinfo {title} {Quantum information
  with {Rydberg} atoms},\ }\href {https://doi.org/10/cbqkmb} {\bibfield
  {journal} {\bibinfo  {journal} {Rev. Mod. Phys.}\ }\textbf {\bibinfo {volume}
  {82}},\ \bibinfo {pages} {2313} (\bibinfo {year} {2010})}\BibitemShut
  {NoStop}%
\bibitem [{\citenamefont {Adams}\ \emph {et~al.}(2019)\citenamefont {Adams},
  \citenamefont {Pritchard},\ and\ \citenamefont {Shaffer}}]{Adams2019}%
  \BibitemOpen
  \bibfield  {author} {\bibinfo {author} {\bibfnamefont {C.~S.}\ \bibnamefont
  {Adams}}, \bibinfo {author} {\bibfnamefont {J.~D.}\ \bibnamefont
  {Pritchard}},\ and\ \bibinfo {author} {\bibfnamefont {J.~P.}\ \bibnamefont
  {Shaffer}},\ }\bibfield  {title} {\bibinfo {title} {Rydberg atom quantum
  technologies},\ }\href {https://doi.org/10.1088/1361-6455/ab52ef} {\bibfield
  {journal} {\bibinfo  {journal} {J. Phys. B: At. Mol. Opt. Phys.}\ }\textbf
  {\bibinfo {volume} {53}},\ \bibinfo {pages} {012002} (\bibinfo {year}
  {2019})}\BibitemShut {NoStop}%
\bibitem [{\citenamefont {Simons}\ \emph {et~al.}(2021)\citenamefont {Simons},
  \citenamefont {{Artusio-Glimpse}}, \citenamefont {Robinson}, \citenamefont
  {Prajapati},\ and\ \citenamefont {Holloway}}]{Simons2021}%
  \BibitemOpen
  \bibfield  {author} {\bibinfo {author} {\bibfnamefont {M.~T.}\ \bibnamefont
  {Simons}}, \bibinfo {author} {\bibfnamefont {A.~B.}\ \bibnamefont
  {{Artusio-Glimpse}}}, \bibinfo {author} {\bibfnamefont {A.~K.}\ \bibnamefont
  {Robinson}}, \bibinfo {author} {\bibfnamefont {N.}~\bibnamefont
  {Prajapati}},\ and\ \bibinfo {author} {\bibfnamefont {C.~L.}\ \bibnamefont
  {Holloway}},\ }\bibfield  {title} {\bibinfo {title} {Rydberg atom-based
  sensors for radio-frequency electric field metrology, sensing, and
  communications},\ }\href {https://doi.org/10.1016/j.measen.2021.100273}
  {\bibfield  {journal} {\bibinfo  {journal} {Measurement: Sensors}\ }\textbf
  {\bibinfo {volume} {18}},\ \bibinfo {pages} {100273} (\bibinfo {year}
  {2021})}\BibitemShut {NoStop}%
\bibitem [{\citenamefont {Osterwalder}\ and\ \citenamefont
  {Merkt}(1999)}]{Osterwalder1999}%
  \BibitemOpen
  \bibfield  {author} {\bibinfo {author} {\bibfnamefont {A.}~\bibnamefont
  {Osterwalder}}\ and\ \bibinfo {author} {\bibfnamefont {F.}~\bibnamefont
  {Merkt}},\ }\bibfield  {title} {\bibinfo {title} {Using {{High Rydberg
  States}} as {{Electric Field Sensors}}},\ }\href
  {https://doi.org/10.1103/PhysRevLett.82.1831} {\bibfield  {journal} {\bibinfo
   {journal} {Phys. Rev. Lett.}\ }\textbf {\bibinfo {volume} {82}},\ \bibinfo
  {pages} {1831} (\bibinfo {year} {1999})}\BibitemShut {NoStop}%
\bibitem [{\citenamefont {Mohapatra}\ \emph {et~al.}(2008)\citenamefont
  {Mohapatra}, \citenamefont {Bason}, \citenamefont {Butscher}, \citenamefont
  {Weatherill},\ and\ \citenamefont {Adams}}]{Mohapatra2008}%
  \BibitemOpen
  \bibfield  {author} {\bibinfo {author} {\bibfnamefont {A.~K.}\ \bibnamefont
  {Mohapatra}}, \bibinfo {author} {\bibfnamefont {M.~G.}\ \bibnamefont
  {Bason}}, \bibinfo {author} {\bibfnamefont {B.}~\bibnamefont {Butscher}},
  \bibinfo {author} {\bibfnamefont {K.~J.}\ \bibnamefont {Weatherill}},\ and\
  \bibinfo {author} {\bibfnamefont {C.~S.}\ \bibnamefont {Adams}},\ }\bibfield
  {title} {\bibinfo {title} {A giant electro-optic effect using polarizable
  dark states},\ }\href@noop {} {\bibfield  {journal} {\bibinfo  {journal}
  {Nature Phys.}\ }\textbf {\bibinfo {volume} {4}},\ \bibinfo {pages} {890}
  (\bibinfo {year} {2008})}\BibitemShut {NoStop}%
\bibitem [{\citenamefont {Sedlacek}\ \emph {et~al.}(2012)\citenamefont
  {Sedlacek}, \citenamefont {Schwettmann}, \citenamefont {K{\"u}bler},
  \citenamefont {L{\"o}w}, \citenamefont {Pfau},\ and\ \citenamefont
  {Shaffer}}]{Sedlacek2012}%
  \BibitemOpen
  \bibfield  {author} {\bibinfo {author} {\bibfnamefont {J.~A.}\ \bibnamefont
  {Sedlacek}}, \bibinfo {author} {\bibfnamefont {A.}~\bibnamefont
  {Schwettmann}}, \bibinfo {author} {\bibfnamefont {H.}~\bibnamefont
  {K{\"u}bler}}, \bibinfo {author} {\bibfnamefont {R.}~\bibnamefont {L{\"o}w}},
  \bibinfo {author} {\bibfnamefont {T.}~\bibnamefont {Pfau}},\ and\ \bibinfo
  {author} {\bibfnamefont {J.~P.}\ \bibnamefont {Shaffer}},\ }\bibfield
  {title} {\bibinfo {title} {Microwave electrometry with {{Rydberg}} atoms in a
  vapour cell using bright atomic resonances},\ }\href
  {https://doi.org/10.1038/nphys2423} {\bibfield  {journal} {\bibinfo
  {journal} {Nature Phys}\ }\textbf {\bibinfo {volume} {8}},\ \bibinfo {pages}
  {819} (\bibinfo {year} {2012})}\BibitemShut {NoStop}%
\bibitem [{\citenamefont {Wade}\ \emph {et~al.}(2017)\citenamefont {Wade},
  \citenamefont {{\v{S}}ibali{\'{c}}}, \citenamefont {de~Melo}, \citenamefont
  {Kondo}, \citenamefont {Adams},\ and\ \citenamefont {Weatherill}}]{Wade2017}%
  \BibitemOpen
  \bibfield  {author} {\bibinfo {author} {\bibfnamefont {C.~G.}\ \bibnamefont
  {Wade}}, \bibinfo {author} {\bibfnamefont {N.}~\bibnamefont
  {{\v{S}}ibali{\'{c}}}}, \bibinfo {author} {\bibfnamefont {N.~R.}\
  \bibnamefont {de~Melo}}, \bibinfo {author} {\bibfnamefont {J.~M.}\
  \bibnamefont {Kondo}}, \bibinfo {author} {\bibfnamefont {C.~S.}\ \bibnamefont
  {Adams}},\ and\ \bibinfo {author} {\bibfnamefont {K.~J.}\ \bibnamefont
  {Weatherill}},\ }\bibfield  {title} {\bibinfo {title} {Real-time near-field
  terahertz imaging with atomic optical fluorescence},\ }\href
  {https://doi.org/10.1038/nphoton.2016.214} {\bibfield  {journal} {\bibinfo
  {journal} {Nat. Photonics}\ }\textbf {\bibinfo {volume} {11}},\ \bibinfo
  {pages} {40} (\bibinfo {year} {2017})}\BibitemShut {NoStop}%
\bibitem [{\citenamefont {Downes}\ \emph {et~al.}(2020)\citenamefont {Downes},
  \citenamefont {MacKellar}, \citenamefont {Whiting}, \citenamefont
  {Bourgenot}, \citenamefont {Adams},\ and\ \citenamefont
  {Weatherill}}]{Downes2020}%
  \BibitemOpen
  \bibfield  {author} {\bibinfo {author} {\bibfnamefont {L.~A.}\ \bibnamefont
  {Downes}}, \bibinfo {author} {\bibfnamefont {A.~R.}\ \bibnamefont
  {MacKellar}}, \bibinfo {author} {\bibfnamefont {D.~J.}\ \bibnamefont
  {Whiting}}, \bibinfo {author} {\bibfnamefont {C.}~\bibnamefont {Bourgenot}},
  \bibinfo {author} {\bibfnamefont {C.~S.}\ \bibnamefont {Adams}},\ and\
  \bibinfo {author} {\bibfnamefont {K.~J.}\ \bibnamefont {Weatherill}},\
  }\bibfield  {title} {\bibinfo {title} {Full-field terahertz imaging at
  kilohertz frame rates using atomic vapor},\ }\href
  {https://doi.org/10.1103/PhysRevX.10.011027} {\bibfield  {journal} {\bibinfo
  {journal} {Phys. Rev. X}\ }\textbf {\bibinfo {volume} {10}},\ \bibinfo
  {pages} {011027} (\bibinfo {year} {2020})}\BibitemShut {NoStop}%
\bibitem [{\citenamefont {Wade}\ \emph {et~al.}(2018)\citenamefont {Wade},
  \citenamefont {Marcuzzi}, \citenamefont {Levi}, \citenamefont {Kondo},
  \citenamefont {Lesanovsky}, \citenamefont {Adams},\ and\ \citenamefont
  {Weatherill}}]{Wade2018}%
  \BibitemOpen
  \bibfield  {author} {\bibinfo {author} {\bibfnamefont {C.~G.}\ \bibnamefont
  {Wade}}, \bibinfo {author} {\bibfnamefont {M.}~\bibnamefont {Marcuzzi}},
  \bibinfo {author} {\bibfnamefont {E.}~\bibnamefont {Levi}}, \bibinfo {author}
  {\bibfnamefont {J.~M.}\ \bibnamefont {Kondo}}, \bibinfo {author}
  {\bibfnamefont {I.}~\bibnamefont {Lesanovsky}}, \bibinfo {author}
  {\bibfnamefont {C.~S.}\ \bibnamefont {Adams}},\ and\ \bibinfo {author}
  {\bibfnamefont {K.~J.}\ \bibnamefont {Weatherill}},\ }\bibfield  {title}
  {\bibinfo {title} {A terahertz-driven non-equilibrium phase transition in a
  room temperature atomic vapour},\ }\href@noop {} {\bibfield  {journal}
  {\bibinfo  {journal} {Nature Commun.}\ }\textbf {\bibinfo {volume} {9}},\
  \bibinfo {pages} {3567} (\bibinfo {year} {2018})}\BibitemShut {NoStop}%
\bibitem [{\citenamefont {Ding}\ \emph {et~al.}(2022)\citenamefont {Ding},
  \citenamefont {Liu}, \citenamefont {Shi}, \citenamefont {Guo}, \citenamefont
  {M{\o}lmer},\ and\ \citenamefont {Adams}}]{Ding2022}%
  \BibitemOpen
  \bibfield  {author} {\bibinfo {author} {\bibfnamefont {D.-S.}\ \bibnamefont
  {Ding}}, \bibinfo {author} {\bibfnamefont {Z.-K.}\ \bibnamefont {Liu}},
  \bibinfo {author} {\bibfnamefont {B.-S.}\ \bibnamefont {Shi}}, \bibinfo
  {author} {\bibfnamefont {G.-C.}\ \bibnamefont {Guo}}, \bibinfo {author}
  {\bibfnamefont {K.}~\bibnamefont {M{\o}lmer}},\ and\ \bibinfo {author}
  {\bibfnamefont {C.~S.}\ \bibnamefont {Adams}},\ }\bibfield  {title} {\bibinfo
  {title} {Enhanced metrology at the critical point of a many-body {{Rydberg}}
  atomic system},\ }\href {https://doi.org/10.1038/s41567-022-01777-8}
  {\bibfield  {journal} {\bibinfo  {journal} {Nature Physics}\ }\textbf
  {\bibinfo {volume} {18}},\ \bibinfo {pages} {1447} (\bibinfo {year}
  {2022})}\BibitemShut {NoStop}%
\bibitem [{\citenamefont {Eckner}\ \emph {et~al.}(2023)\citenamefont {Eckner},
  \citenamefont {Darkwah~Oppong}, \citenamefont {Cao}, \citenamefont {Young},
  \citenamefont {Milner}, \citenamefont {Robinson}, \citenamefont {Ye},\ and\
  \citenamefont {Kaufman}}]{Eckner2023}%
  \BibitemOpen
  \bibfield  {author} {\bibinfo {author} {\bibfnamefont {W.~J.}\ \bibnamefont
  {Eckner}}, \bibinfo {author} {\bibfnamefont {N.}~\bibnamefont
  {Darkwah~Oppong}}, \bibinfo {author} {\bibfnamefont {A.}~\bibnamefont {Cao}},
  \bibinfo {author} {\bibfnamefont {A.~W.}\ \bibnamefont {Young}}, \bibinfo
  {author} {\bibfnamefont {W.~R.}\ \bibnamefont {Milner}}, \bibinfo {author}
  {\bibfnamefont {J.~M.}\ \bibnamefont {Robinson}}, \bibinfo {author}
  {\bibfnamefont {J.}~\bibnamefont {Ye}},\ and\ \bibinfo {author}
  {\bibfnamefont {A.~M.}\ \bibnamefont {Kaufman}},\ }\bibfield  {title}
  {\bibinfo {title} {Realizing spin squeezing with {{Rydberg}} interactions in
  an optical clock},\ }\href {https://doi.org/10.1038/s41586-023-06360-6}
  {\bibfield  {journal} {\bibinfo  {journal} {Nature}\ }\textbf {\bibinfo
  {volume} {621}},\ \bibinfo {pages} {734} (\bibinfo {year}
  {2023})}\BibitemShut {NoStop}%
\bibitem [{\citenamefont {Hines}\ \emph {et~al.}(2023)\citenamefont {Hines},
  \citenamefont {Rajagopal}, \citenamefont {Moreau}, \citenamefont {Wahrman},
  \citenamefont {Lewis}, \citenamefont {Markovi{\'c}},\ and\ \citenamefont
  {{Schleier-Smith}}}]{Hines2023}%
  \BibitemOpen
  \bibfield  {author} {\bibinfo {author} {\bibfnamefont {J.~A.}\ \bibnamefont
  {Hines}}, \bibinfo {author} {\bibfnamefont {S.~V.}\ \bibnamefont
  {Rajagopal}}, \bibinfo {author} {\bibfnamefont {G.~L.}\ \bibnamefont
  {Moreau}}, \bibinfo {author} {\bibfnamefont {M.~D.}\ \bibnamefont {Wahrman}},
  \bibinfo {author} {\bibfnamefont {N.~A.}\ \bibnamefont {Lewis}}, \bibinfo
  {author} {\bibfnamefont {O.}~\bibnamefont {Markovi{\'c}}},\ and\ \bibinfo
  {author} {\bibfnamefont {M.}~\bibnamefont {{Schleier-Smith}}},\ }\bibfield
  {title} {\bibinfo {title} {Spin {{Squeezing}} by {{Rydberg Dressing}} in an
  {{Array}} of {{Atomic Ensembles}}},\ }\href
  {https://doi.org/10.1103/PhysRevLett.131.063401} {\bibfield  {journal}
  {\bibinfo  {journal} {Phys. Rev. Lett.}\ }\textbf {\bibinfo {volume} {131}},\
  \bibinfo {pages} {063401} (\bibinfo {year} {2023})}\BibitemShut {NoStop}%
\bibitem [{\citenamefont {Yamamoto}\ and\ \citenamefont
  {Matsuki}(1999)}]{Yamamoto1999}%
  \BibitemOpen
  \bibfield  {author} {\bibinfo {author} {\bibfnamefont {K.}~\bibnamefont
  {Yamamoto}}\ and\ \bibinfo {author} {\bibfnamefont {S.}~\bibnamefont
  {Matsuki}},\ }\bibfield  {title} {\bibinfo {title} {Quantum analysis of the
  rydberg atom cavity detector of dark matter axions},\ }\href
  {https://doi.org/https://doi.org/10.1016/S0920-5632(98)00515-5} {\bibfield
  {journal} {\bibinfo  {journal} {Nuclear Physics B - Proceedings Supplements}\
  }\textbf {\bibinfo {volume} {72}},\ \bibinfo {pages} {132} (\bibinfo {year}
  {1999})},\ \bibinfo {note} {proceedings of the 5th IFT Workshop on
  Axions}\BibitemShut {NoStop}%
\bibitem [{\citenamefont {Graham}\ \emph {et~al.}(2023)\citenamefont {Graham},
  \citenamefont {Ghosh}, \citenamefont {Zhu}, \citenamefont {Bai},
  \citenamefont {Cahn}, \citenamefont {Durcan}, \citenamefont {Jewell},
  \citenamefont {Speller}, \citenamefont {Zacarias}, \citenamefont {Zhou},\
  and\ \citenamefont {Maruyama}}]{Graham2023}%
  \BibitemOpen
  \bibfield  {author} {\bibinfo {author} {\bibfnamefont {E.}~\bibnamefont
  {Graham}}, \bibinfo {author} {\bibfnamefont {S.}~\bibnamefont {Ghosh}},
  \bibinfo {author} {\bibfnamefont {Y.}~\bibnamefont {Zhu}}, \bibinfo {author}
  {\bibfnamefont {X.}~\bibnamefont {Bai}}, \bibinfo {author} {\bibfnamefont
  {S.~B.}\ \bibnamefont {Cahn}}, \bibinfo {author} {\bibfnamefont
  {E.}~\bibnamefont {Durcan}}, \bibinfo {author} {\bibfnamefont {M.~J.}\
  \bibnamefont {Jewell}}, \bibinfo {author} {\bibfnamefont {D.~H.}\
  \bibnamefont {Speller}}, \bibinfo {author} {\bibfnamefont {S.~M.}\
  \bibnamefont {Zacarias}}, \bibinfo {author} {\bibfnamefont {L.~T.}\
  \bibnamefont {Zhou}},\ and\ \bibinfo {author} {\bibfnamefont {R.~H.}\
  \bibnamefont {Maruyama}},\ }\bibfield  {title} {\bibinfo {title}
  {Rydberg-atom-based single-photon detection for haloscope axion searches},\
  }\bibfield  {journal} {\bibinfo  {journal} {arXiv}\ }\href
  {https://doi.org/10.48550/arXiv.2310.15352} {10.48550/arXiv.2310.15352}
  (\bibinfo {year} {2023})\BibitemShut {NoStop}%
\bibitem [{\citenamefont {Browaeys}\ and\ \citenamefont
  {Lahaye}(2020)}]{Browaeys2020}%
  \BibitemOpen
  \bibfield  {author} {\bibinfo {author} {\bibfnamefont {A.}~\bibnamefont
  {Browaeys}}\ and\ \bibinfo {author} {\bibfnamefont {T.}~\bibnamefont
  {Lahaye}},\ }\bibfield  {title} {\bibinfo {title} {Many-body physics with
  individually controlled {Rydberg} atoms},\ }\href
  {https://doi.org/10.1038/s41567-019-0733-z} {\bibfield  {journal} {\bibinfo
  {journal} {Nat. Phys.}\ }\textbf {\bibinfo {volume} {16}},\ \bibinfo {pages}
  {132} (\bibinfo {year} {2020})}\BibitemShut {NoStop}%
\bibitem [{\citenamefont {Amthor}\ \emph {et~al.}(2010)\citenamefont {Amthor},
  \citenamefont {Giese}, \citenamefont {Hofmann},\ and\ \citenamefont
  {Weidemüller}}]{Amthor2010}%
  \BibitemOpen
  \bibfield  {author} {\bibinfo {author} {\bibfnamefont {T.}~\bibnamefont
  {Amthor}}, \bibinfo {author} {\bibfnamefont {C.}~\bibnamefont {Giese}},
  \bibinfo {author} {\bibfnamefont {C.~S.}\ \bibnamefont {Hofmann}},\ and\
  \bibinfo {author} {\bibfnamefont {M.}~\bibnamefont {Weidemüller}},\
  }\bibfield  {title} {\bibinfo {title} {Evidence of {A}ntiblockade in an
  ultracold {R}ydberg gas},\ }\href
  {https://doi.org/10.1103/physrevlett.104.013001} {\bibfield  {journal}
  {\bibinfo  {journal} {Phys. Rev. Lett.}\ }\textbf {\bibinfo {volume} {104}},\
  \bibinfo {pages} {013001} (\bibinfo {year} {2010})}\BibitemShut {NoStop}%
\bibitem [{\citenamefont {Ates}\ \emph {et~al.}(2007)\citenamefont {Ates},
  \citenamefont {Pohl}, \citenamefont {Pattard},\ and\ \citenamefont
  {Rost}}]{Ates2007}%
  \BibitemOpen
  \bibfield  {author} {\bibinfo {author} {\bibfnamefont {C.}~\bibnamefont
  {Ates}}, \bibinfo {author} {\bibfnamefont {T.}~\bibnamefont {Pohl}}, \bibinfo
  {author} {\bibfnamefont {T.}~\bibnamefont {Pattard}},\ and\ \bibinfo {author}
  {\bibfnamefont {J.~M.}\ \bibnamefont {Rost}},\ }\bibfield  {title} {\bibinfo
  {title} {{A}ntiblockade in {R}ydberg excitation of an ultracold lattice
  gas},\ }\href {https://doi.org/10.1103/physrevlett.98.023002} {\bibfield
  {journal} {\bibinfo  {journal} {Phys. Rev. Lett.}\ }\textbf {\bibinfo
  {volume} {98}},\ \bibinfo {pages} {023002} (\bibinfo {year}
  {2007})}\BibitemShut {NoStop}%
\bibitem [{\citenamefont {Mattioli}\ \emph {et~al.}(2015)\citenamefont
  {Mattioli}, \citenamefont {Glätzle},\ and\ \citenamefont
  {Lechner}}]{Mattioli2015}%
  \BibitemOpen
  \bibfield  {author} {\bibinfo {author} {\bibfnamefont {M.}~\bibnamefont
  {Mattioli}}, \bibinfo {author} {\bibfnamefont {A.~W.}\ \bibnamefont
  {Glätzle}},\ and\ \bibinfo {author} {\bibfnamefont {W.}~\bibnamefont
  {Lechner}},\ }\bibfield  {title} {\bibinfo {title} {{From classical to
  quantum non-equilibrium dynamics of Rydberg excitations in optical
  lattices}},\ }\href {https://doi.org/10.1088/1367-2630/17/11/113039}
  {\bibfield  {journal} {\bibinfo  {journal} {New J. Phys.}\ }\textbf {\bibinfo
  {volume} {17}},\ \bibinfo {pages} {113039} (\bibinfo {year}
  {2015})}\BibitemShut {NoStop}%
\bibitem [{\citenamefont {Valado}\ \emph {et~al.}(2016)\citenamefont {Valado},
  \citenamefont {Simonelli}, \citenamefont {Hoogerland}, \citenamefont
  {Lesanovsky}, \citenamefont {Garrahan}, \citenamefont {Arimondo},
  \citenamefont {Ciampini},\ and\ \citenamefont {Morsch}}]{Valado2016}%
  \BibitemOpen
  \bibfield  {author} {\bibinfo {author} {\bibfnamefont {M.~M.}\ \bibnamefont
  {Valado}}, \bibinfo {author} {\bibfnamefont {C.}~\bibnamefont {Simonelli}},
  \bibinfo {author} {\bibfnamefont {M.~D.}\ \bibnamefont {Hoogerland}},
  \bibinfo {author} {\bibfnamefont {I.}~\bibnamefont {Lesanovsky}}, \bibinfo
  {author} {\bibfnamefont {J.~P.}\ \bibnamefont {Garrahan}}, \bibinfo {author}
  {\bibfnamefont {E.}~\bibnamefont {Arimondo}}, \bibinfo {author}
  {\bibfnamefont {D.}~\bibnamefont {Ciampini}},\ and\ \bibinfo {author}
  {\bibfnamefont {O.}~\bibnamefont {Morsch}},\ }\bibfield  {title} {\bibinfo
  {title} {Experimental observation of controllable kinetic constraints in a
  cold atomic gas},\ }\href {https://doi.org/10.1103/PhysRevA.93.040701}
  {\bibfield  {journal} {\bibinfo  {journal} {Phys. Rev. A}\ }\textbf {\bibinfo
  {volume} {93}},\ \bibinfo {pages} {040701} (\bibinfo {year}
  {2016})}\BibitemShut {NoStop}%
\bibitem [{\citenamefont {Letscher}\ \emph {et~al.}(2017)\citenamefont
  {Letscher}, \citenamefont {Thomas}, \citenamefont {Niederpr\"um},
  \citenamefont {Fleischhauer},\ and\ \citenamefont {Ott}}]{Letscher2017}%
  \BibitemOpen
  \bibfield  {author} {\bibinfo {author} {\bibfnamefont {F.}~\bibnamefont
  {Letscher}}, \bibinfo {author} {\bibfnamefont {O.}~\bibnamefont {Thomas}},
  \bibinfo {author} {\bibfnamefont {T.}~\bibnamefont {Niederpr\"um}}, \bibinfo
  {author} {\bibfnamefont {M.}~\bibnamefont {Fleischhauer}},\ and\ \bibinfo
  {author} {\bibfnamefont {H.}~\bibnamefont {Ott}},\ }\bibfield  {title}
  {\bibinfo {title} {{Bistability Versus Metastability in Driven Dissipative
  Rydberg Gases}},\ }\href {https://doi.org/10.1103/PhysRevX.7.021020}
  {\bibfield  {journal} {\bibinfo  {journal} {Phys. Rev. X}\ }\textbf {\bibinfo
  {volume} {7}},\ \bibinfo {pages} {021020} (\bibinfo {year}
  {2017})}\BibitemShut {NoStop}%
\bibitem [{\citenamefont {Su}\ \emph {et~al.}(2017)\citenamefont {Su},
  \citenamefont {Gao}, \citenamefont {Liang},\ and\ \citenamefont
  {Zhang}}]{Su2017}%
  \BibitemOpen
  \bibfield  {author} {\bibinfo {author} {\bibfnamefont {S.-L.}\ \bibnamefont
  {Su}}, \bibinfo {author} {\bibfnamefont {Y.}~\bibnamefont {Gao}}, \bibinfo
  {author} {\bibfnamefont {E.}~\bibnamefont {Liang}},\ and\ \bibinfo {author}
  {\bibfnamefont {S.}~\bibnamefont {Zhang}},\ }\bibfield  {title} {\bibinfo
  {title} {{Fast Rydberg antiblockade regime and its applications in quantum
  logic gates}},\ }\href {https://doi.org/10.1103/PhysRevA.95.022319}
  {\bibfield  {journal} {\bibinfo  {journal} {Phys. Rev. A}\ }\textbf {\bibinfo
  {volume} {95}},\ \bibinfo {pages} {022319} (\bibinfo {year}
  {2017})}\BibitemShut {NoStop}%
\bibitem [{\citenamefont {Gambetta}\ \emph {et~al.}(2020)\citenamefont
  {Gambetta}, \citenamefont {Zhang}, \citenamefont {Hennrich}, \citenamefont
  {Lesanovsky},\ and\ \citenamefont {Li}}]{Gambetta2020a}%
  \BibitemOpen
  \bibfield  {author} {\bibinfo {author} {\bibfnamefont {F.~M.}\ \bibnamefont
  {Gambetta}}, \bibinfo {author} {\bibfnamefont {C.}~\bibnamefont {Zhang}},
  \bibinfo {author} {\bibfnamefont {M.}~\bibnamefont {Hennrich}}, \bibinfo
  {author} {\bibfnamefont {I.}~\bibnamefont {Lesanovsky}},\ and\ \bibinfo
  {author} {\bibfnamefont {W.}~\bibnamefont {Li}},\ }\bibfield  {title}
  {\bibinfo {title} {Long-range multibody interactions and three-body
  {A}ntiblockade in a trapped {R}ydberg ion chain},\ }\href
  {https://doi.org/10.1103/PhysRevLett.125.133602} {\bibfield  {journal}
  {\bibinfo  {journal} {Phys. Rev. Lett.}\ }\textbf {\bibinfo {volume} {125}},\
  \bibinfo {pages} {133602} (\bibinfo {year} {2020})}\BibitemShut {NoStop}%
\bibitem [{\citenamefont {Festa}\ \emph {et~al.}(2022)\citenamefont {Festa},
  \citenamefont {Lorenz}, \citenamefont {Steinert}, \citenamefont {Chen},
  \citenamefont {Osterholz}, \citenamefont {Eberhard},\ and\ \citenamefont
  {Gross}}]{Festa2022}%
  \BibitemOpen
  \bibfield  {author} {\bibinfo {author} {\bibfnamefont {L.}~\bibnamefont
  {Festa}}, \bibinfo {author} {\bibfnamefont {N.}~\bibnamefont {Lorenz}},
  \bibinfo {author} {\bibfnamefont {L.-M.}\ \bibnamefont {Steinert}}, \bibinfo
  {author} {\bibfnamefont {Z.}~\bibnamefont {Chen}}, \bibinfo {author}
  {\bibfnamefont {P.}~\bibnamefont {Osterholz}}, \bibinfo {author}
  {\bibfnamefont {R.}~\bibnamefont {Eberhard}},\ and\ \bibinfo {author}
  {\bibfnamefont {C.}~\bibnamefont {Gross}},\ }\bibfield  {title} {\bibinfo
  {title} {Blackbody-radiation-induced facilitated excitation of {R}ydberg
  atoms in optical tweezers},\ }\href
  {https://doi.org/10.1103/physreva.105.013109} {\bibfield  {journal} {\bibinfo
   {journal} {Phys. Rev. A}\ }\textbf {\bibinfo {volume} {105}},\ \bibinfo
  {pages} {013109} (\bibinfo {year} {2022})}\BibitemShut {NoStop}%
\bibitem [{\citenamefont {Liu}\ \emph {et~al.}(2022)\citenamefont {Liu},
  \citenamefont {Yang}, \citenamefont {Bienias}, \citenamefont {Iadecola},\
  and\ \citenamefont {Gorshkov}}]{Liu2022}%
  \BibitemOpen
  \bibfield  {author} {\bibinfo {author} {\bibfnamefont {F.}~\bibnamefont
  {Liu}}, \bibinfo {author} {\bibfnamefont {Z.-C.}\ \bibnamefont {Yang}},
  \bibinfo {author} {\bibfnamefont {P.}~\bibnamefont {Bienias}}, \bibinfo
  {author} {\bibfnamefont {T.}~\bibnamefont {Iadecola}},\ and\ \bibinfo
  {author} {\bibfnamefont {A.~V.}\ \bibnamefont {Gorshkov}},\ }\bibfield
  {title} {\bibinfo {title} {{Localization and Criticality in Antiblockaded
  Two-Dimensional Rydberg Atom Arrays}},\ }\href
  {https://doi.org/10.1103/PhysRevLett.128.013603} {\bibfield  {journal}
  {\bibinfo  {journal} {Phys. Rev. Lett.}\ }\textbf {\bibinfo {volume} {128}},\
  \bibinfo {pages} {013603} (\bibinfo {year} {2022})}\BibitemShut {NoStop}%
\bibitem [{\citenamefont {Gross}\ and\ \citenamefont
  {Haroche}(1982)}]{Gross1982}%
  \BibitemOpen
  \bibfield  {author} {\bibinfo {author} {\bibfnamefont {M.}~\bibnamefont
  {Gross}}\ and\ \bibinfo {author} {\bibfnamefont {S.}~\bibnamefont
  {Haroche}},\ }\bibfield  {title} {\bibinfo {title} {Superradiance: An essay
  on the theory of collective spontaneous emission},\ }\href
  {https://doi.org/https://doi.org/10.1016/0370-1573(82)90102-8} {\bibfield
  {journal} {\bibinfo  {journal} {Phys. Rep.}\ }\textbf {\bibinfo {volume}
  {93}},\ \bibinfo {pages} {301} (\bibinfo {year} {1982})}\BibitemShut
  {NoStop}%
\bibitem [{\citenamefont {Hao}\ \emph {et~al.}(2021)\citenamefont {Hao},
  \citenamefont {Bai}, \citenamefont {Bai}, \citenamefont {Bai}, \citenamefont
  {Jiao}, \citenamefont {Huang}, \citenamefont {Zhao}, \citenamefont {Li},\
  and\ \citenamefont {Jia}}]{Hao2021}%
  \BibitemOpen
  \bibfield  {author} {\bibinfo {author} {\bibfnamefont {L.}~\bibnamefont
  {Hao}}, \bibinfo {author} {\bibfnamefont {Z.}~\bibnamefont {Bai}}, \bibinfo
  {author} {\bibfnamefont {J.}~\bibnamefont {Bai}}, \bibinfo {author}
  {\bibfnamefont {S.}~\bibnamefont {Bai}}, \bibinfo {author} {\bibfnamefont
  {Y.}~\bibnamefont {Jiao}}, \bibinfo {author} {\bibfnamefont {G.}~\bibnamefont
  {Huang}}, \bibinfo {author} {\bibfnamefont {J.}~\bibnamefont {Zhao}},
  \bibinfo {author} {\bibfnamefont {W.}~\bibnamefont {Li}},\ and\ \bibinfo
  {author} {\bibfnamefont {S.}~\bibnamefont {Jia}},\ }\bibfield  {title}
  {\bibinfo {title} {Observation of blackbody radiation enhanced superradiance
  in ultracold {R}ydberg gases},\ }\href
  {https://doi.org/10.1088/1367-2630/ac136c} {\bibfield  {journal} {\bibinfo
  {journal} {New J. Phys.}\ }\textbf {\bibinfo {volume} {23}},\ \bibinfo
  {pages} {083017} (\bibinfo {year} {2021})}\BibitemShut {NoStop}%
\bibitem [{\citenamefont {Suarez}\ \emph {et~al.}(2022)\citenamefont {Suarez},
  \citenamefont {Wolf}, \citenamefont {Weiss},\ and\ \citenamefont
  {Slama}}]{Suarez2022}%
  \BibitemOpen
  \bibfield  {author} {\bibinfo {author} {\bibfnamefont {E.}~\bibnamefont
  {Suarez}}, \bibinfo {author} {\bibfnamefont {P.}~\bibnamefont {Wolf}},
  \bibinfo {author} {\bibfnamefont {P.}~\bibnamefont {Weiss}},\ and\ \bibinfo
  {author} {\bibfnamefont {S.}~\bibnamefont {Slama}},\ }\bibfield  {title}
  {\bibinfo {title} {Superradiance decoherence caused by long-range
  {{Rydberg-atom}} pair interactions},\ }\href
  {https://doi.org/10.1103/PhysRevA.105.L041302} {\bibfield  {journal}
  {\bibinfo  {journal} {Phys. Rev. A}\ }\textbf {\bibinfo {volume} {105}},\
  \bibinfo {pages} {L041302} (\bibinfo {year} {2022})}\BibitemShut {NoStop}%
\bibitem [{sup()}]{supplement}%
  \BibitemOpen
  \href@noop {} {}\bibinfo {note} {See the Supplemental Material, which further
  contains Refs.~\cite{Wiseman2009,Johansson2012,Johansson2013}, for
  details.}\BibitemShut {Stop}%
\bibitem [{\citenamefont {Johansson}\ \emph {et~al.}(2012)\citenamefont
  {Johansson}, \citenamefont {Nation},\ and\ \citenamefont
  {Nori}}]{Johansson2012}%
  \BibitemOpen
  \bibfield  {author} {\bibinfo {author} {\bibfnamefont {J.~R.}\ \bibnamefont
  {Johansson}}, \bibinfo {author} {\bibfnamefont {P.~D.}\ \bibnamefont
  {Nation}},\ and\ \bibinfo {author} {\bibfnamefont {F.}~\bibnamefont {Nori}},\
  }\bibfield  {title} {\bibinfo {title} {{QuTiP}: An open-source {Python}
  framework for the dynamics of open quantum systems},\ }\href
  {https://doi.org/10.1016/j.cpc.2012.02.021} {\bibfield  {journal} {\bibinfo
  {journal} {Computer Physics Communications}\ }\textbf {\bibinfo {volume}
  {183}},\ \bibinfo {pages} {1760} (\bibinfo {year} {2012})}\BibitemShut
  {NoStop}%
\bibitem [{\citenamefont {Johansson}\ \emph {et~al.}(2013)\citenamefont
  {Johansson}, \citenamefont {Nation},\ and\ \citenamefont
  {Nori}}]{Johansson2013}%
  \BibitemOpen
  \bibfield  {author} {\bibinfo {author} {\bibfnamefont {J.~R.}\ \bibnamefont
  {Johansson}}, \bibinfo {author} {\bibfnamefont {P.~D.}\ \bibnamefont
  {Nation}},\ and\ \bibinfo {author} {\bibfnamefont {F.}~\bibnamefont {Nori}},\
  }\bibfield  {title} {\bibinfo {title} {{QuTiP} 2: A {Python} framework for
  the dynamics of open quantum systems},\ }\href
  {https://doi.org/10.1016/j.cpc.2012.11.019} {\bibfield  {journal} {\bibinfo
  {journal} {Computer Physics Communications}\ }\textbf {\bibinfo {volume}
  {184}},\ \bibinfo {pages} {1234} (\bibinfo {year} {2013})}\BibitemShut
  {NoStop}%
\bibitem [{\citenamefont {Bettles}\ \emph {et~al.}(2016)\citenamefont
  {Bettles}, \citenamefont {Gardiner},\ and\ \citenamefont
  {Adams}}]{Bettles2016}%
  \BibitemOpen
  \bibfield  {author} {\bibinfo {author} {\bibfnamefont {R.~J.}\ \bibnamefont
  {Bettles}}, \bibinfo {author} {\bibfnamefont {S.~A.}\ \bibnamefont
  {Gardiner}},\ and\ \bibinfo {author} {\bibfnamefont {C.~S.}\ \bibnamefont
  {Adams}},\ }\bibfield  {title} {\bibinfo {title} {Enhanced optical cross
  section via collective coupling of atomic dipoles in a 2d array},\ }\href
  {https://doi.org/10.1103/PhysRevLett.116.103602} {\bibfield  {journal}
  {\bibinfo  {journal} {Phys. Rev. Lett.}\ }\textbf {\bibinfo {volume} {116}},\
  \bibinfo {pages} {103602} (\bibinfo {year} {2016})}\BibitemShut {NoStop}%
\bibitem [{\citenamefont {Weber}\ \emph {et~al.}(2017)\citenamefont {Weber},
  \citenamefont {Tresp}, \citenamefont {Menke}, \citenamefont {Urvoy},
  \citenamefont {Firstenberg}, \citenamefont {B{\"u}chler},\ and\ \citenamefont
  {Hofferberth}}]{Weber2017}%
  \BibitemOpen
  \bibfield  {author} {\bibinfo {author} {\bibfnamefont {S.}~\bibnamefont
  {Weber}}, \bibinfo {author} {\bibfnamefont {C.}~\bibnamefont {Tresp}},
  \bibinfo {author} {\bibfnamefont {H.}~\bibnamefont {Menke}}, \bibinfo
  {author} {\bibfnamefont {A.}~\bibnamefont {Urvoy}}, \bibinfo {author}
  {\bibfnamefont {O.}~\bibnamefont {Firstenberg}}, \bibinfo {author}
  {\bibfnamefont {H.~P.}\ \bibnamefont {B{\"u}chler}},\ and\ \bibinfo {author}
  {\bibfnamefont {S.}~\bibnamefont {Hofferberth}},\ }\bibfield  {title}
  {\bibinfo {title} {{Tutorial: Calculation of Rydberg interaction
  potentials}},\ }\href {https://doi.org/10.1088/1361-6455/aa743a} {\bibfield
  {journal} {\bibinfo  {journal} {J. Phys. B: At. Mol. Opt. Phys.}\ }\textbf
  {\bibinfo {volume} {50}},\ \bibinfo {pages} {133001} (\bibinfo {year}
  {2017})}\BibitemShut {NoStop}%
\bibitem [{\citenamefont {Marcuzzi}\ \emph {et~al.}(2017)\citenamefont
  {Marcuzzi}, \citenamefont {Minar}, \citenamefont {Barredo}, \citenamefont
  {de~L\'es\'eleuc}, \citenamefont {Labuhn}, \citenamefont {Lahaye},
  \citenamefont {Browaeys}, \citenamefont {Levi},\ and\ \citenamefont
  {Lesanovsky}}]{Marcuzzi2017}%
  \BibitemOpen
  \bibfield  {author} {\bibinfo {author} {\bibfnamefont {M.}~\bibnamefont
  {Marcuzzi}}, \bibinfo {author} {\bibfnamefont {J.}~\bibnamefont {Minar}},
  \bibinfo {author} {\bibfnamefont {D.}~\bibnamefont {Barredo}}, \bibinfo
  {author} {\bibfnamefont {S.}~\bibnamefont {de~L\'es\'eleuc}}, \bibinfo
  {author} {\bibfnamefont {H.}~\bibnamefont {Labuhn}}, \bibinfo {author}
  {\bibfnamefont {T.}~\bibnamefont {Lahaye}}, \bibinfo {author} {\bibfnamefont
  {A.}~\bibnamefont {Browaeys}}, \bibinfo {author} {\bibfnamefont
  {E.}~\bibnamefont {Levi}},\ and\ \bibinfo {author} {\bibfnamefont
  {I.}~\bibnamefont {Lesanovsky}},\ }\bibfield  {title} {\bibinfo {title}
  {Facilitation dynamics and localization phenomena in {Rydberg} lattice gases
  with position disorder},\ }\href
  {https://doi.org/10.1103/PhysRevLett.118.063606} {\bibfield  {journal}
  {\bibinfo  {journal} {Phys. Rev. Lett.}\ }\textbf {\bibinfo {volume} {118}},\
  \bibinfo {pages} {063606} (\bibinfo {year} {2017})}\BibitemShut {NoStop}%
\bibitem [{\citenamefont {Ostmann}\ \emph {et~al.}(2019)\citenamefont
  {Ostmann}, \citenamefont {Marcuzzi}, \citenamefont {Garrahan},\ and\
  \citenamefont {Lesanovsky}}]{Ostmann2019}%
  \BibitemOpen
  \bibfield  {author} {\bibinfo {author} {\bibfnamefont {M.}~\bibnamefont
  {Ostmann}}, \bibinfo {author} {\bibfnamefont {M.}~\bibnamefont {Marcuzzi}},
  \bibinfo {author} {\bibfnamefont {J.~P.}\ \bibnamefont {Garrahan}},\ and\
  \bibinfo {author} {\bibfnamefont {I.}~\bibnamefont {Lesanovsky}},\ }\bibfield
   {title} {\bibinfo {title} {Localization in spin chains with facilitation
  constraints and disordered interactions},\ }\href
  {https://doi.org/10.1103/PhysRevA.99.060101} {\bibfield  {journal} {\bibinfo
  {journal} {Phys. Rev. A}\ }\textbf {\bibinfo {volume} {99}},\ \bibinfo
  {pages} {060101} (\bibinfo {year} {2019})}\BibitemShut {NoStop}%
\bibitem [{\citenamefont {Wintermantel}\ \emph {et~al.}(2021)\citenamefont
  {Wintermantel}, \citenamefont {Buchhold}, \citenamefont {Shevate},
  \citenamefont {Morgado}, \citenamefont {Wang}, \citenamefont {Lochead},
  \citenamefont {Diehl},\ and\ \citenamefont {Whitlock}}]{Wintermantel2021}%
  \BibitemOpen
  \bibfield  {author} {\bibinfo {author} {\bibfnamefont {T.}~\bibnamefont
  {Wintermantel}}, \bibinfo {author} {\bibfnamefont {M.}~\bibnamefont
  {Buchhold}}, \bibinfo {author} {\bibfnamefont {S.}~\bibnamefont {Shevate}},
  \bibinfo {author} {\bibfnamefont {M.}~\bibnamefont {Morgado}}, \bibinfo
  {author} {\bibfnamefont {Y.}~\bibnamefont {Wang}}, \bibinfo {author}
  {\bibfnamefont {G.}~\bibnamefont {Lochead}}, \bibinfo {author} {\bibfnamefont
  {S.}~\bibnamefont {Diehl}},\ and\ \bibinfo {author} {\bibfnamefont
  {S.}~\bibnamefont {Whitlock}},\ }\bibfield  {title} {\bibinfo {title}
  {Epidemic growth and griffiths effects on an emergent network of excited
  atoms},\ }\href {https://doi.org/https://doi.org/10.1038/s41467-020-20333-7}
  {\bibfield  {journal} {\bibinfo  {journal} {Nature Commun.}\ }\textbf
  {\bibinfo {volume} {12}},\ \bibinfo {pages} {103} (\bibinfo {year}
  {2021})}\BibitemShut {NoStop}%
\bibitem [{\citenamefont {Anderson}\ \emph {et~al.}(2011)\citenamefont
  {Anderson}, \citenamefont {Younge},\ and\ \citenamefont
  {Raithel}}]{Anderson2011}%
  \BibitemOpen
  \bibfield  {author} {\bibinfo {author} {\bibfnamefont {S.~E.}\ \bibnamefont
  {Anderson}}, \bibinfo {author} {\bibfnamefont {K.~C.}\ \bibnamefont
  {Younge}},\ and\ \bibinfo {author} {\bibfnamefont {G.}~\bibnamefont
  {Raithel}},\ }\bibfield  {title} {\bibinfo {title} {Trapping rydberg atoms in
  an optical lattice},\ }\href {https://doi.org/10/fxwcmn} {\bibfield
  {journal} {\bibinfo  {journal} {Phys. Rev. Lett.}\ }\textbf {\bibinfo
  {volume} {107}},\ \bibinfo {pages} {263001} (\bibinfo {year}
  {2011})}\BibitemShut {NoStop}%
\bibitem [{\citenamefont {Li}\ \emph {et~al.}(2013)\citenamefont {Li},
  \citenamefont {Dudin},\ and\ \citenamefont {Kuzmich}}]{Li2013}%
  \BibitemOpen
  \bibfield  {author} {\bibinfo {author} {\bibfnamefont {L.}~\bibnamefont
  {Li}}, \bibinfo {author} {\bibfnamefont {Y.~O.}\ \bibnamefont {Dudin}},\ and\
  \bibinfo {author} {\bibfnamefont {A.}~\bibnamefont {Kuzmich}},\ }\bibfield
  {title} {\bibinfo {title} {Entanglement between light and an optical atomic
  excitation},\ }\href {https://doi.org/10/f4283f} {\bibfield  {journal}
  {\bibinfo  {journal} {Nature}\ }\textbf {\bibinfo {volume} {498}},\ \bibinfo
  {pages} {466} (\bibinfo {year} {2013})}\BibitemShut {NoStop}%
\bibitem [{\citenamefont {Goldschmidt}\ \emph {et~al.}(2015)\citenamefont
  {Goldschmidt}, \citenamefont {Norris}, \citenamefont {Koller}, \citenamefont
  {Wyllie}, \citenamefont {Brown}, \citenamefont {Porto}, \citenamefont
  {Safronova},\ and\ \citenamefont {Safronova}}]{Goldschmidt2015}%
  \BibitemOpen
  \bibfield  {author} {\bibinfo {author} {\bibfnamefont {E.~A.}\ \bibnamefont
  {Goldschmidt}}, \bibinfo {author} {\bibfnamefont {D.~G.}\ \bibnamefont
  {Norris}}, \bibinfo {author} {\bibfnamefont {S.~B.}\ \bibnamefont {Koller}},
  \bibinfo {author} {\bibfnamefont {R.}~\bibnamefont {Wyllie}}, \bibinfo
  {author} {\bibfnamefont {R.~C.}\ \bibnamefont {Brown}}, \bibinfo {author}
  {\bibfnamefont {J.~V.}\ \bibnamefont {Porto}}, \bibinfo {author}
  {\bibfnamefont {U.~I.}\ \bibnamefont {Safronova}},\ and\ \bibinfo {author}
  {\bibfnamefont {M.~S.}\ \bibnamefont {Safronova}},\ }\bibfield  {title}
  {\bibinfo {title} {Magic wavelengths for the 5s\text{--}{1}{8}s transition in
  {R}ubidium},\ }\href {https://doi.org/10/gnzsd8} {\bibfield  {journal}
  {\bibinfo  {journal} {Phys. Rev. A}\ }\textbf {\bibinfo {volume} {91}},\
  \bibinfo {pages} {032518} (\bibinfo {year} {2015})}\BibitemShut {NoStop}%
\bibitem [{\citenamefont {Wilson}\ \emph {et~al.}(2022)\citenamefont {Wilson},
  \citenamefont {Saskin}, \citenamefont {Meng}, \citenamefont {Ma},
  \citenamefont {Dilip}, \citenamefont {Burgers},\ and\ \citenamefont
  {Thompson}}]{Wilson2022}%
  \BibitemOpen
  \bibfield  {author} {\bibinfo {author} {\bibfnamefont {J.~T.}\ \bibnamefont
  {Wilson}}, \bibinfo {author} {\bibfnamefont {S.}~\bibnamefont {Saskin}},
  \bibinfo {author} {\bibfnamefont {Y.}~\bibnamefont {Meng}}, \bibinfo {author}
  {\bibfnamefont {S.}~\bibnamefont {Ma}}, \bibinfo {author} {\bibfnamefont
  {R.}~\bibnamefont {Dilip}}, \bibinfo {author} {\bibfnamefont {A.~P.}\
  \bibnamefont {Burgers}},\ and\ \bibinfo {author} {\bibfnamefont {J.~D.}\
  \bibnamefont {Thompson}},\ }\bibfield  {title} {\bibinfo {title} {Trapping
  alkaline earth {R}ydberg atoms optical tweezer arrays},\ }\href
  {https://doi.org/10.1103/PhysRevLett.128.033201} {\bibfield  {journal}
  {\bibinfo  {journal} {Phys. Rev. Lett.}\ }\textbf {\bibinfo {volume} {128}},\
  \bibinfo {pages} {033201} (\bibinfo {year} {2022})}\BibitemShut {NoStop}%
\bibitem [{\citenamefont {Kaufman}\ \emph {et~al.}(2012)\citenamefont
  {Kaufman}, \citenamefont {Lester},\ and\ \citenamefont
  {Regal}}]{Kaufman2012}%
  \BibitemOpen
  \bibfield  {author} {\bibinfo {author} {\bibfnamefont {A.~M.}\ \bibnamefont
  {Kaufman}}, \bibinfo {author} {\bibfnamefont {B.~J.}\ \bibnamefont
  {Lester}},\ and\ \bibinfo {author} {\bibfnamefont {C.~A.}\ \bibnamefont
  {Regal}},\ }\bibfield  {title} {\bibinfo {title} {Cooling a single atom in an
  optical tweezer to its quantum ground state},\ }\href
  {https://doi.org/10/gddjnd} {\bibfield  {journal} {\bibinfo  {journal} {Phys.
  Rev. X}\ }\textbf {\bibinfo {volume} {2}},\ \bibinfo {pages} {041014}
  (\bibinfo {year} {2012})}\BibitemShut {NoStop}%
\bibitem [{\citenamefont {Thompson}\ \emph {et~al.}(2013)\citenamefont
  {Thompson}, \citenamefont {Tiecke}, \citenamefont {Zibrov}, \citenamefont
  {Vuletić},\ and\ \citenamefont {Lukin}}]{Thompson2013}%
  \BibitemOpen
  \bibfield  {author} {\bibinfo {author} {\bibfnamefont {J.~D.}\ \bibnamefont
  {Thompson}}, \bibinfo {author} {\bibfnamefont {T.~G.}\ \bibnamefont
  {Tiecke}}, \bibinfo {author} {\bibfnamefont {A.~S.}\ \bibnamefont {Zibrov}},
  \bibinfo {author} {\bibfnamefont {V.}~\bibnamefont {Vuletić}},\ and\
  \bibinfo {author} {\bibfnamefont {M.~D.}\ \bibnamefont {Lukin}},\ }\bibfield
  {title} {\bibinfo {title} {Coherence and {Raman} sideband cooling of a single
  atom in an optical tweezer},\ }\href
  {https://doi.org/10.1103/PhysRevLett.110.133001} {\bibfield  {journal}
  {\bibinfo  {journal} {Phys. Rev. Lett.}\ }\textbf {\bibinfo {volume} {110}},\
  \bibinfo {pages} {133001} (\bibinfo {year} {2013})}\BibitemShut {NoStop}%
\bibitem [{\citenamefont {Cooper}\ \emph {et~al.}(2018)\citenamefont {Cooper},
  \citenamefont {Covey}, \citenamefont {Madjarov}, \citenamefont {Porsev},
  \citenamefont {Safronova},\ and\ \citenamefont {Endres}}]{Cooper2018}%
  \BibitemOpen
  \bibfield  {author} {\bibinfo {author} {\bibfnamefont {A.}~\bibnamefont
  {Cooper}}, \bibinfo {author} {\bibfnamefont {J.~P.}\ \bibnamefont {Covey}},
  \bibinfo {author} {\bibfnamefont {I.~S.}\ \bibnamefont {Madjarov}}, \bibinfo
  {author} {\bibfnamefont {S.~G.}\ \bibnamefont {Porsev}}, \bibinfo {author}
  {\bibfnamefont {M.~S.}\ \bibnamefont {Safronova}},\ and\ \bibinfo {author}
  {\bibfnamefont {M.}~\bibnamefont {Endres}},\ }\bibfield  {title} {\bibinfo
  {title} {Alkaline-earth atoms in optical tweezers},\ }\href
  {https://doi.org/10/gfvs72} {\bibfield  {journal} {\bibinfo  {journal} {Phys.
  Rev. X}\ }\textbf {\bibinfo {volume} {8}},\ \bibinfo {pages} {041055}
  (\bibinfo {year} {2018})}\BibitemShut {NoStop}%
\bibitem [{\citenamefont {Norcia}\ \emph {et~al.}(2018)\citenamefont {Norcia},
  \citenamefont {Young},\ and\ \citenamefont {Kaufman}}]{Norcia2018}%
  \BibitemOpen
  \bibfield  {author} {\bibinfo {author} {\bibfnamefont {M.}~\bibnamefont
  {Norcia}}, \bibinfo {author} {\bibfnamefont {A.}~\bibnamefont {Young}},\ and\
  \bibinfo {author} {\bibfnamefont {A.}~\bibnamefont {Kaufman}},\ }\bibfield
  {title} {\bibinfo {title} {Microscopic control and detection of ultracold
  strontium in optical-tweezer arrays},\ }\href {https://doi.org/10/gfvs74}
  {\bibfield  {journal} {\bibinfo  {journal} {Phys. Rev. X}\ }\textbf {\bibinfo
  {volume} {8}},\ \bibinfo {pages} {041054} (\bibinfo {year}
  {2018})}\BibitemShut {NoStop}%
\bibitem [{\citenamefont {Saskin}\ \emph {et~al.}(2019)\citenamefont {Saskin},
  \citenamefont {Wilson}, \citenamefont {Grinkemeyer},\ and\ \citenamefont
  {Thompson}}]{Saskin2019}%
  \BibitemOpen
  \bibfield  {author} {\bibinfo {author} {\bibfnamefont {S.}~\bibnamefont
  {Saskin}}, \bibinfo {author} {\bibfnamefont {J.~T.}\ \bibnamefont {Wilson}},
  \bibinfo {author} {\bibfnamefont {B.}~\bibnamefont {Grinkemeyer}},\ and\
  \bibinfo {author} {\bibfnamefont {J.~D.}\ \bibnamefont {Thompson}},\
  }\bibfield  {title} {\bibinfo {title} {Narrow-{{Line Cooling}} and
  {{Imaging}} of {{Ytterbium Atoms}} in an {{Optical Tweezer Array}}},\ }\href
  {https://doi.org/10/gfzbt5} {\bibfield  {journal} {\bibinfo  {journal} {Phys.
  Rev. Lett.}\ }\textbf {\bibinfo {volume} {122}},\ \bibinfo {pages} {143002}
  (\bibinfo {year} {2019})}\BibitemShut {NoStop}%
\bibitem [{\citenamefont {Lorenz}\ \emph {et~al.}(2021)\citenamefont {Lorenz},
  \citenamefont {Festa}, \citenamefont {Steinert},\ and\ \citenamefont
  {Gross}}]{Lorenz2021}%
  \BibitemOpen
  \bibfield  {author} {\bibinfo {author} {\bibfnamefont {N.}~\bibnamefont
  {Lorenz}}, \bibinfo {author} {\bibfnamefont {L.}~\bibnamefont {Festa}},
  \bibinfo {author} {\bibfnamefont {L.-M.}\ \bibnamefont {Steinert}},\ and\
  \bibinfo {author} {\bibfnamefont {C.}~\bibnamefont {Gross}},\ }\bibfield
  {title} {\bibinfo {title} {Raman sideband cooling in optical tweezer arrays
  for {{Rydberg}} dressing},\ }\href
  {https://doi.org/10.21468/scipostphys.10.3.052} {\bibfield  {journal}
  {\bibinfo  {journal} {SciPost Phys.}\ }\textbf {\bibinfo {volume} {10}},\
  \bibinfo {pages} {052} (\bibinfo {year} {2021})}\BibitemShut {NoStop}%
\bibitem [{\citenamefont {Hölzl}\ \emph {et~al.}(2023)\citenamefont {Hölzl},
  \citenamefont {Götzelmann}, \citenamefont {Wirth}, \citenamefont
  {Safronova}, \citenamefont {Weber},\ and\ \citenamefont
  {Meinert}}]{Hoelzl2023}%
  \BibitemOpen
  \bibfield  {author} {\bibinfo {author} {\bibfnamefont {C.}~\bibnamefont
  {Hölzl}}, \bibinfo {author} {\bibfnamefont {A.}~\bibnamefont {Götzelmann}},
  \bibinfo {author} {\bibfnamefont {M.}~\bibnamefont {Wirth}}, \bibinfo
  {author} {\bibfnamefont {M.~S.}\ \bibnamefont {Safronova}}, \bibinfo {author}
  {\bibfnamefont {S.}~\bibnamefont {Weber}},\ and\ \bibinfo {author}
  {\bibfnamefont {F.}~\bibnamefont {Meinert}},\ }\bibfield  {title} {\bibinfo
  {title} {Motional ground-state cooling of single atoms in state-dependent
  optical tweezers},\ }\href {https://doi.org/10.1103/PhysRevResearch.5.033093}
  {\bibfield  {journal} {\bibinfo  {journal} {Phys. Rev. Res.}\ }\textbf
  {\bibinfo {volume} {5}},\ \bibinfo {pages} {033093} (\bibinfo {year}
  {2023})}\BibitemShut {NoStop}%
\bibitem [{\citenamefont {Magoni}\ \emph {et~al.}(2022)\citenamefont {Magoni},
  \citenamefont {Mazza},\ and\ \citenamefont {Lesanovsky}}]{Magoni2022}%
  \BibitemOpen
  \bibfield  {author} {\bibinfo {author} {\bibfnamefont {M.}~\bibnamefont
  {Magoni}}, \bibinfo {author} {\bibfnamefont {P.}~\bibnamefont {Mazza}},\ and\
  \bibinfo {author} {\bibfnamefont {I.}~\bibnamefont {Lesanovsky}},\ }\bibfield
   {title} {\bibinfo {title} {Phonon dressing of a facilitated one-dimensional
  {R}ydberg lattice gas},\ }\href {https://doi.org/10/gr7nh5} {\bibfield
  {journal} {\bibinfo  {journal} {SciPost Phys. Core}\ }\textbf {\bibinfo
  {volume} {5}},\ \bibinfo {pages} {041} (\bibinfo {year} {2022})}\BibitemShut
  {NoStop}%
\bibitem [{\citenamefont {Magoni}\ \emph {et~al.}(2023)\citenamefont {Magoni},
  \citenamefont {Nill},\ and\ \citenamefont {Lesanovsky}}]{Magoni2023a}%
  \BibitemOpen
  \bibfield  {author} {\bibinfo {author} {\bibfnamefont {M.}~\bibnamefont
  {Magoni}}, \bibinfo {author} {\bibfnamefont {C.}~\bibnamefont {Nill}},\ and\
  \bibinfo {author} {\bibfnamefont {I.}~\bibnamefont {Lesanovsky}},\ }\bibfield
   {title} {\bibinfo {title} {Coherent spin-phonon scattering in facilitated
  {Rydberg} lattices},\ }\bibfield  {journal} {\bibinfo  {journal} {arXiv}\
  }\href {https://doi.org/10.48550/arXiv.2311.00064}
  {10.48550/arXiv.2311.00064} (\bibinfo {year} {2023})\BibitemShut {NoStop}%
\bibitem [{\citenamefont {Vidal}(2003)}]{Vidal2003}%
  \BibitemOpen
  \bibfield  {author} {\bibinfo {author} {\bibfnamefont {G.}~\bibnamefont
  {Vidal}},\ }\bibfield  {title} {\bibinfo {title} {Efficient classical
  simulation of slightly entangled quantum computations},\ }\href
  {https://doi.org/10.1103/PhysRevLett.91.147902} {\bibfield  {journal}
  {\bibinfo  {journal} {Phys. Rev. Lett.}\ }\textbf {\bibinfo {volume} {91}},\
  \bibinfo {pages} {147902} (\bibinfo {year} {2003})}\BibitemShut {NoStop}%
\bibitem [{\citenamefont {Vidal}(2004)}]{Vidal2004}%
  \BibitemOpen
  \bibfield  {author} {\bibinfo {author} {\bibfnamefont {G.}~\bibnamefont
  {Vidal}},\ }\bibfield  {title} {\bibinfo {title} {Efficient simulation of
  one-dimensional quantum many-body systems},\ }\href
  {https://doi.org/10.1103/PhysRevLett.93.040502} {\bibfield  {journal}
  {\bibinfo  {journal} {Phys. Rev. Lett.}\ }\textbf {\bibinfo {volume} {93}},\
  \bibinfo {pages} {040502} (\bibinfo {year} {2004})}\BibitemShut {NoStop}%
\bibitem [{\citenamefont {Or{\'{u}}s}(2014)}]{Orus2014}%
  \BibitemOpen
  \bibfield  {author} {\bibinfo {author} {\bibfnamefont {R.}~\bibnamefont
  {Or{\'{u}}s}},\ }\bibfield  {title} {\bibinfo {title} {A practical
  introduction to tensor networks: Matrix product states and projected
  entangled pair states},\ }\href {https://doi.org/10.1016/j.aop.2014.06.013}
  {\bibfield  {journal} {\bibinfo  {journal} {Ann. Phys. (NY)}\ }\textbf
  {\bibinfo {volume} {349}},\ \bibinfo {pages} {117} (\bibinfo {year}
  {2014})}\BibitemShut {NoStop}%
\bibitem [{\citenamefont {Paeckel}\ \emph {et~al.}(2019)\citenamefont
  {Paeckel}, \citenamefont {Köhler}, \citenamefont {Swoboda}, \citenamefont
  {Manmana}, \citenamefont {Schollwöck},\ and\ \citenamefont
  {Hubig}}]{Paeckel2019}%
  \BibitemOpen
  \bibfield  {author} {\bibinfo {author} {\bibfnamefont {S.}~\bibnamefont
  {Paeckel}}, \bibinfo {author} {\bibfnamefont {T.}~\bibnamefont {Köhler}},
  \bibinfo {author} {\bibfnamefont {A.}~\bibnamefont {Swoboda}}, \bibinfo
  {author} {\bibfnamefont {S.~R.}\ \bibnamefont {Manmana}}, \bibinfo {author}
  {\bibfnamefont {U.}~\bibnamefont {Schollwöck}},\ and\ \bibinfo {author}
  {\bibfnamefont {C.}~\bibnamefont {Hubig}},\ }\bibfield  {title} {\bibinfo
  {title} {Time-evolution methods for matrix-product states},\ }\href
  {https://doi.org/https://doi.org/10.1016/j.aop.2019.167998} {\bibfield
  {journal} {\bibinfo  {journal} {Ann. Phys. (NY)}\ }\textbf {\bibinfo {volume}
  {411}},\ \bibinfo {pages} {167998} (\bibinfo {year} {2019})}\BibitemShut
  {NoStop}%
\bibitem [{\citenamefont {Cirac}\ \emph {et~al.}(2021)\citenamefont {Cirac},
  \citenamefont {P{\'{e}}rez-Garc{\'{\i}}a}, \citenamefont {Schuch},\ and\
  \citenamefont {Verstraete}}]{Cirac2021}%
  \BibitemOpen
  \bibfield  {author} {\bibinfo {author} {\bibfnamefont {J.~I.}\ \bibnamefont
  {Cirac}}, \bibinfo {author} {\bibfnamefont {D.}~\bibnamefont
  {P{\'{e}}rez-Garc{\'{\i}}a}}, \bibinfo {author} {\bibfnamefont
  {N.}~\bibnamefont {Schuch}},\ and\ \bibinfo {author} {\bibfnamefont
  {F.}~\bibnamefont {Verstraete}},\ }\bibfield  {title} {\bibinfo {title}
  {Matrix product states and projected entangled pair states: Concepts,
  symmetries, theorems},\ }\href {https://doi.org/10.1103/revmodphys.93.045003}
  {\bibfield  {journal} {\bibinfo  {journal} {Rev. Mod. Phys.}\ }\textbf
  {\bibinfo {volume} {93}},\ \bibinfo {pages} {045003} (\bibinfo {year}
  {2021})}\BibitemShut {NoStop}%
\bibitem [{\citenamefont {Gray}(2018)}]{Gray2018}%
  \BibitemOpen
  \bibfield  {author} {\bibinfo {author} {\bibfnamefont {J.}~\bibnamefont
  {Gray}},\ }\bibfield  {title} {\bibinfo {title} {quimb: A {Python} package
  for quantum information and many-body calculations},\ }\href
  {https://doi.org/10.21105/joss.00819} {\bibfield  {journal} {\bibinfo
  {journal} {Journal of Open Source Software}\ }\textbf {\bibinfo {volume}
  {3}},\ \bibinfo {pages} {819} (\bibinfo {year} {2018})}\BibitemShut {NoStop}%
\bibitem [{\citenamefont {Wiseman}\ and\ \citenamefont
  {Milburn}(2009)}]{Wiseman2009}%
  \BibitemOpen
  \bibfield  {author} {\bibinfo {author} {\bibfnamefont {H.~M.}\ \bibnamefont
  {Wiseman}}\ and\ \bibinfo {author} {\bibfnamefont {G.~J.}\ \bibnamefont
  {Milburn}},\ }\href@noop {} {\emph {\bibinfo {title} {Quantum measurement and
  control}}}\ (\bibinfo  {publisher} {Cambridge university press},\ \bibinfo
  {year} {2009})\BibitemShut {NoStop}%
\end{thebibliography}%

\setcounter{equation}{0}
\setcounter{figure}{0}
\setcounter{table}{0}
\makeatletter
\renewcommand{\theequation}{S\arabic{equation}}
\renewcommand{\thefigure}{S\arabic{figure}}

\makeatletter
\renewcommand{\theequation}{S\arabic{equation}}
\renewcommand{\thefigure}{S\arabic{figure}}

%%%%%%%%%%%%%%%%%%%%%%%%%%%%%%%%%%%%%%%%%%%%%%%%%%%%%%%%%%%%%%%
%% COMMENT THIS OUT FOR PHYS REV VERSION
%%
%% YOU NEED IT IF YOU WOULD LIKE TO COMPILE
%% THE APPENDIX TO THE SAME PDF FOR ARXIV VERSION

\clearpage

\onecolumngrid
\setcounter{page}{1}

\begin{center}
{\Large SUPPLEMENTAL MATERIAL}
\end{center}
\begin{center}
\vspace{0.8cm}
{\Large Avalanche terahertz photon detection in a Rydberg tweezer array}
\end{center}
\begin{center}
Chris Nill$^{1}$, Albert Cabot$^{1}$, Arno Trautmann$^{2}$, Christian Groß$^{2}$, and Igor Lesanovsky$^{1}$
\end{center}
\begin{center}
$^1${\em Institut f\"ur Theoretische Physik, Universit\"at T\"ubingen,}\\
{\em Auf der Morgenstelle 14, 72076 T\"ubingen, Germany}\\
$^2${\em Physikalisches Institut, Universit\"{a}t T\"{u}bingen, Auf der Morgenstelle 14, 72076 T\"{u}bingen, Germany}
\end{center}

\section{Description of the absorption process}
In this section we write down a master equation for the density matrix of the system $\rho$. This can describe incoherent processes occurring in the system such as the absorption of the THz photon, the emission of light by the atoms or dephasing processes. We then discuss in which conditions we can approximate the dynamics as unitary, as we have done in the main text.

\subsection{Local absorption}
The Markovian master equation describing the incoherent processes occurring in the system is given by
\begin{equation}
\partial_t \rho=-i[H,\rho]+\sum_j \mathcal{D}[L_j]\rho,  
\end{equation}
where  $\mathcal{D}[L_j]\rho=L_j\rho L_j^\dagger-\{L_j^\dagger L_j,\rho\}/2$ is the standard Lindblad dissipator, $L_j$ is the jump operator describing the specific incoherent process and $H$ is the Hamiltonian of the system \cite{Wiseman2009}. In the case of modeling the local absorption of a THz photon, the jump operator is $L_j=\sqrt{\Gamma_\mathrm{THz}}\dyad{r}{e}_j$. This leads to the following master equation:
\begin{equation}
\partial_t \rho=-i[H,\rho]+\Gamma_\mathrm{THz}\sum_{j=1}^N \mathcal{D}[\dyad{r}{e}_j]\rho.
\end{equation}

In our setting, it is  relevant to focus on individual experimental realizations rather than on the average description provided by the master equation. Thus, we consider the quantum trajectory unraveling of the master equation \cite{Wiseman2009}. In this approach, the dynamics of the system is described by a stochastic process for a pure state (the conditioned state of the system) rather than for the density matrix. This stochastic process consists of a continuous time evolution through a non-Hermitian effective Hamiltonian interceded with sudden jumps, that model the absorption of a photon. The periods in between  photon absorptions can be modeled as
\begin{equation}
\dv{t}\ket*{\tilde{\Psi}(t)}=-iH_\mathrm{eff}\ket*{\tilde{\Psi}(t)},
\end{equation}
where
\begin{equation}
H_\mathrm{eff}=H-i\frac{\Gamma_\mathrm{THz}}{2}\sum_{j=1}^N n_j^\mathrm{(e)}    
\end{equation}
is the effective Hamiltonian and we use a tilde to denote an unnormalized state. Right after site $k$ absorbs a photon, the state of the system changes to:
\begin{equation}
|\Psi(t+dt)\rangle= \mathcal{N}\dyad{r}{e}_k|\tilde{\Psi}(t)\rangle,   
\end{equation}
where $\mathcal{N}$ is a normalization constant. The probability of occurrence of the absorption of a photon is contained in the norm of $|\tilde{\Psi}(t)\rangle$ \cite{Wiseman2009}. Notice that the application of this jump operator onto the state $|\Psi_\mathrm{s}\rangle$ leads to $|\Psi_\mathrm{er}\rangle$ as discussed in the main text. Moreover, if $\Gamma_\mathrm{THz}T_\mathrm{s}\ll1$, the dynamics in between the absorption of photons can be approximated as unitary, as the effects of the non-Hermitian part of the effective Hamiltonian will be small.

\subsection{Collective absorption}
In the case of a collective absorption, the jump operator is $L=\sqrt{\Gamma_\mathrm{THz}}\sum_{j=1}^N \dyad{r}{e}_j$. This leads to the following master equation describing the evolution of the density matrix:
\begin{equation}
\partial_t \rho=-i[H,\rho]+\Gamma_\mathrm{THz} \mathcal{D}\left[\sum_{j=1}^N \dyad{r}{e}_j\right]\rho.    
\end{equation}

In the quantum trajectory approach, the effective Hamiltonian changes to:
\begin{equation}
H^\mathrm{c}_\mathrm{eff}=H-i\frac{\Gamma_\mathrm{THz}}{2}\Bigg(\sum_{j=1}^N \dyad{e}{r}_j\Bigg)\Bigg(\sum_{k=1}^N \dyad{r}{e}_k\Bigg).    
\end{equation}
The state of the system after a collective absorption changes according to:
\begin{equation}
\ket{\Psi(t+\mathrm{d}t)}= \mathcal{N}\Bigg(\sum_{j=1}^N \dyad{r}{e}_j\Bigg)\ket*{\tilde{\Psi}(t)}.   
\end{equation}
where $\mathcal{N}$ is a normalization constant. Notice that the application of this jump operator onto the state $|\Psi_\mathrm{s}\rangle$ yields $|\Psi_\mathrm{er}^\mathrm{c}\rangle$ as discussed in the main text. Hence, in this case, the Rydberg excitation is collective over the whole array. Again, a unitary description of the dynamics in between photon absorptions is possible when $\Gamma_\mathrm{THz}T_\mathrm{s}\ll1$. Finally,  it is interesting to note that for the initial state $|\Psi_\mathrm{s}\rangle$ the rate of absorption induced by the local or the collective process is the same, since $\sum_j\langle \Psi_\mathrm{s}|L^\dagger_j L_j|\Psi_\mathrm{s}\rangle =\langle \Psi_\mathrm{s}|L^\dagger L|\Psi_\mathrm{s}\rangle$. Nevertheless, once the first photon has been absorbed, the collective process displays an enhanced absorption rate with respect to the local one, due to the correlations present in $|\Psi^\mathrm{c}_\mathrm{er}\rangle$.

\section{Rabi frequency determines amplification velocity}
\begin{figure}[h]
    \centering
    \includegraphics{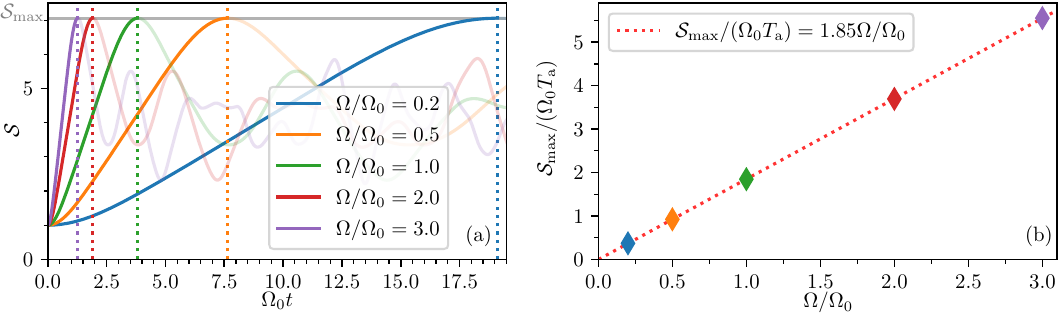}
    \caption{\textbf{Proportionality of the amplification velocity and the Rabi frequency.}
    (a) Time evolution of the signal $\mathcal{S}$ after a local absorption of a THz photon.
    The time $T_\mathrm{a}$ required to reach the maximum signal $\mathcal{S}_\mathrm{max}$ varies for different Rabi frequencies $\Omega$ and is marked by the dashed vertical lines.
    The Rabi frequency is normalized to the Rabi frequency used in Fig. 2 of the main paper, denoted $\Omega_0$. For the simulation, $V_\mathrm{rr}=-\Delta=500 \Omega_0$ was used.
    A higher (lower) Rabi frequency $\Omega$ leads to a faster (slower) attainment of the maximum of signal $\mathcal{S}$.
    The finite-size dynamics, which are shown with opacity, are not considered in the following calculations. 
    (b) Proportionality of the Rabi frequency $\Omega$ and the amplification velocity.
    The diamonds correspond to the Rabi frequency of the same color as in panel (a).
    A proportional relationship with $\mathcal{S}/(\Omega_0 T_\mathrm{a})=1.85\Omega/\Omega_0$ can be observed. Consequently, doubling the Rabi frequency $\Omega$ halves the amplification time $T_\mathrm{a}$.
    The simulation was done integrating the corresponding master equation  \cite{Johansson2012,Johansson2013} of an 1D chain with open boundary conditions and $N=11$ atoms.
    }
    \label{fig:omega_scaling}
\end{figure}

In this section, we explore how a varying Rabi frequency affects the amplification process.
To do so, we conduct simulations using different values of the Rabi frequency $\Omega$ as seen in Figure 2b of the main paper.
We normalize these values by the Rabi frequency $\Omega_0=\Omega_\mathrm{gr}$ used in the main paper to determine their impact.
In Figure \ref{fig:omega_scaling}a, we show the time evolution of the signal $\mathcal{S}$ and demonstrate that a higher (lower) Rabi frequency shortens (lengthens) the time $\tau$ needed to reach the maximum signal $\mathcal{S}_\mathrm{max}$ of the amplification process.
It should be noted that the change in the Rabi frequency has no effect on the value of $\mathcal{S}_\mathrm{max}$.
Figure \ref{fig:omega_scaling}b shows the signal increase per amplification time $\mathcal{S}_\mathrm{max}/(\Omega_0 T_\mathrm{a})$ for the Rabi frequencies of panel \ref{fig:omega_scaling}a. It can be seen that this quantity, which can be understood as the amplification velocity, is proportional to the Rabi frequency. Therefore, increasing the Rabi frequency during the amplification phase leads to a faster amplification. On the other hand, an increased Rabi frequency increases the dark count rate, such that this is a parameter that can be used to optimize the detector for a concrete use case.

\section{Dephasing during amplification mode}
\begin{figure}[h]
    \centering
    \includegraphics{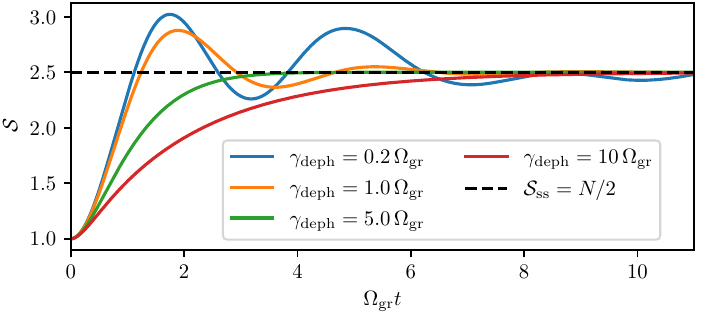}
    \caption{\textbf{Signal amplification including dephasing.} For dephasing rates $\gamma_\mathrm{deph}$ smaller than the Rabi frequency $\Omega_\mathrm{gr}$, finite size oscillations around the steady state $\mathcal{S}_\mathrm{ss}$ are present.
    For $\gamma_\mathrm{deph}\gg\Omega_\mathrm{gr}$ those oscillations vanish and the signal increases asymptotically to the steady state.
    For a short amplification time $T_\mathrm{a}$ at highest signal, a small dephasing is beneficial, while the amplification is still possible even for high dephasing values.
    The simulation was done integrating the corresponding master equation  \cite{Johansson2012,Johansson2013} of an open 1D chain with $N=5$ atoms.
    }
    \label{fig:phase-noise}
\end{figure}

During the amplification mode of the detector, dephasing effects can occur as, e.g., due to the laser phase noise.
In the following we demonstrate how this effect impacts onto the signal amplification process.
Therefore, we model the phase noise on an atom at site $j$ as a jump operator of the form
\begin{equation}
    L^{(j)}_\mathrm{deph}=\sqrt{\gamma_\mathrm{deph}}\dyad{r}_j,
\end{equation}
where $\gamma_\mathrm{deph}$ quantifies the rate of the dephasing.
We calculate the Lindblad time evolution under the Hamiltonian Eq. (1) of the main text including the jump operators on all sites $L^{(j)}_\mathrm{deph}$.
Figure \ref{fig:phase-noise} shows the time evolution of the resulting signal $\mathcal{S}$ for different dephasing rates. We identify a steady state of the signal at $\mathcal{S}_\mathrm{ss}=N/2$. For strong dephasing $(\gamma_\mathrm{deph}\gg\Omega_\mathrm{gr})$ the signal increases asymptotically towards the steady state. In this case, the amplification time $T_\mathrm{a}$ could be extended to increase the amplification factor.
For moderate and small dephasing, finite size effects can be observed in the first Rabi oscillations which decrease with higher dephasing rate.
Consequently, fine-tuning the optimal value for $T_\mathrm{a}$ to the maximum can enable higher values than $N/2$ for the signal $\mathcal{S}$ in the low dephasing regime.

%%%%%%%%%%%%%%%%%%%%%%%%%%%%%%%%%%%%%%%%%%%%%%%%%%%%%%%%%%%%%%%
%% COMMENT THIS OUT FOR ARXIV VERSION
%% YOU NEED IT IF YOU WOULD LIKE TO COMPILE THE APPENDIX ONLY

\clearpage
%\bibliography{biblio}
%%%%%%%%%%%%%%%%%%%%%%%%%%%%%%%%%%%%%%%%%%%%%%%%%%%%%%%%%%%%%%%

%%%%%%%%%%%%%%%%%%%%%%%%%%%%%%%%%%%%%%%%%%%%%%%%%%%%%%%%%%%%%%%

\end{document}